\documentclass[11pt, oneside]{article} 
\usepackage[utf8]{inputenc}
\usepackage[utf8]{inputenc}
\usepackage{geometry}                		
\usepackage{graphicx}				
\usepackage{amssymb}
\usepackage{amsmath}
\usepackage{subfig}
\usepackage{caption}
\usepackage{comment}
\usepackage{amsthm,verbatim,amsfonts,amscd,float,physics}
\usepackage{appendix}

\usepackage[utf8]{inputenc}
\usepackage{xcolor}
\numberwithin{equation}{section}
\usepackage{titling, lipsum}
\title{Quantum Computing and the Riemann Hypothesis}
\author{ Michael McGuigan\\
 email contact: michael.d.mcguigan@gmail.com
}
\date{}

\begin{document}
\begin{titlingpage}

\maketitle
\begin{abstract}
Quantum computing is a promising new area of computing with quantum algorithms offering a potential speedup over classical algorithms if fault tolerant quantum computers can be built. One of the first applications of the classical computer was to the study of the  Riemann hypothesis and quantum computers may be applied to this problem as well. In this paper  we apply the Quantum Fourier Transform (QFT) to study three functions with non-trivial zeros obeying a version of the Riemann hypothesis. We perform our quantum computations with six qubits, but more qubits can be used if quantum error correction allows the QFT algorithm to scale. We represent these functions as ground state wave functions transformed to momentum space. We show how to obtain these functions  as  states in supersymmetric quantum mechanics. Finally we discuss the relation of these functions to $(p,1)$ Random two matrix models at large $N$.

\end{abstract}
\end{titlingpage}

\section{Introduction}

The application of quantum computing to the Riemann hypothesis has been discussed in \cite{vanDam}
\cite{Latorre:2013wvj}
\cite{Sierra:2016rgn}
\cite{Sierra:2014wua} and there have been experiments using ion traps to accurately determining the first few Riemann zeroes \cite{SierraR}
\cite{He}
\cite{He1}
\cite{He2}.
In addition there have been suggestions for the application of quantum computing to the Riemann Hypothesis generated from Artificial Intelligence \cite{gilkalai}. As technology evolves it is interesting to apply it to outstanding challenges to see if new insight can be obtained. The application of quantum computing is especially interesting as the Hilbert-Polya conjecture may provide a link between quantum mechanics and the Riemann hypothesis and quantum computing provides a computational framework which is inherently quantum mechanical. Also the representation of functions and matrices in terms of qubits promises to yield greater computational power over classical computers if fault tolerant machines can be built. 

This paper is organized as follows. In section two we discuss three functions with non-trivial zeros associated with the Fourier transform of a function with no zeros. We represent this in physical terms representing the functions as quantum states in a momentum basis and discuss a deformation of the states which removes the zeros. In section three we discuss how to represent the Quantum Fourier Transform of the states which in principle can be scaled to large values of the three functions. In section four we discuss the representation of the states associated to the three functions to  supersymmetric quantum mechanics. In section five we show how to represent the functions in terms of the $(p,1)$ Random two matrix model and show how this is connected with supersymmetric quantum mechanics. In section six we discuss the main conclusions of the paper.

\section{Three examples of functions with non-trivial zeros}

In \cite {tau} Tau discusses functions that have trivial zeros, no zeros and non-trivial zeros such as the Gamma, Inverse Gamma, trigonometric functions, Riemann zeta function and Riemann xi function. In this section we discuss three functions that have non-trivial zeros and seem to obey analogs of the Riemann hypothesis. The basic method we use to compute these functions is the Fourier transform given by:
\begin{equation}\tilde \psi (p) = \frac{1}{{\sqrt {2\pi } }}\int_{ - \infty }^\infty  {dx{e^{ - ipx}}\psi (x)} \end{equation}
This approach will also give us an opportunity to apply quantum computing to study these functions as the Quantum Fourier transform can be exponentially faster than a standard Fast Fourier transform on a classical computer, assuming a fault tolerant quantum computer can be built. The increase in computer power of quantum computers may allow the study of these functions higher  up on their critical lines than classical computers. Also the framework of quantum computers may be helpful in studying the analogs of the Riemann hypothesis through the Hilbert-Polya conjecture as there is an apparent connection between quantum mechanics which underlies quantum computing and the Riemann zeros.

\subsection*{First function}
The first function we consider is generalized Airy function $Ai_3(p)$ defined through the sum of two Hypergeometric functions as:
\begin{equation}\psi _\varepsilon ^{(1)}(p) =\frac{\, _0F_2\left(;\frac{1}{2},\frac{3}{4};(p+i \epsilon )^4\right)}{\Gamma \left(\frac{3}{4}\right)}-\frac{2 (p+i \epsilon )^2 \, _0F_2\left(;\frac{5}{4},\frac{3}{2};(p+i \epsilon )^4\right)}{\Gamma \left(\frac{5}{4}\right)}\end{equation}
It can be expressed in terms of Fourier integral as:
\begin{equation}\tilde \psi _\varepsilon ^{(1)}(p) = \frac{1}{{\sqrt {2\pi } }}\int_{ - \infty }^\infty  {dx{e^{ - ipx}}\psi _\varepsilon ^{(1)}(x)}  = \frac{1}{{\sqrt {2\pi } }}\int_{ - \infty }^\infty  {dx{e^{ - ipx}}{e^{ - \frac{{{x^4}}}{{256}}}}{e^{ - \varepsilon x}}} \end{equation}
and is related to the Pearcey integral with applications to quantum waves \cite{Lopez}\cite{Ramos}\cite{Lopez2}, black hole physics \cite{Fidkowski:2003nf} and Matrix models. We will discuss the relation to Matrix models in more detail in section 5.

\subsection*{Second function}

The second function we study is a function of the complex order of a K Bessel function. It can be expressed as a Fourier integral as:
\begin{equation}\psi _\varepsilon ^{(2)}(p) = {K_{ip + \varepsilon }}(1) =
\frac{1}{{\sqrt {2\pi } }}\int_{ - \infty }^\infty  {dx{e^{ - ipx}}{e^{ - \cosh x}}{e^{ - \varepsilon x}}} \end{equation}
This function has applications as the wave function of a particle in Rindler space-time.

\subsection*{Third function}

The third function we consider is the Riemann xi function $\xi (ip + \frac{1}{2} + \varepsilon )$ defined by:
\begin{equation}\xi (s) = \frac{1}{2}s(s - 1){\pi ^{ - s/2}}\Gamma (s/2)\zeta (s)\end{equation}
which is an entire function with nontrivial zeros of the zeta function. It can be expressed as Fourier integral through:
\begin{equation}\psi _\varepsilon ^{(3)}(p) = \xi (ip + \frac{1}{2} + \varepsilon ) =
\frac{1}{{\sqrt {2\pi } }}\int_{ - \infty }^\infty  {dx{e^{ - ipx}}\Phi (x){e^{ - \varepsilon x}}} \end{equation}
where the Phi function defined as \cite{Rogers}
\begin{equation}{\Phi  }(x) = \sum\limits_{n = 1}^\infty  {\left( {2{\pi ^2}{n^4}{e^{9x/2}} - 3\pi {n^2}{e^{5x/2}}} \right)} \exp ( - \pi {n^2}{e^{2x}})\end{equation}
the Riemann xi function has many applications to Number theory through its representation as a product involving the prime numbers.

The properties of the three functions are listed in table 1 and many features of the three functions are illustrated in figures 1-6. The most interesting features involve the deformation parameter $\epsilon$. When $\epsilon$ is nonzero the magnitude of the function lifted off the $p$ axis removing the nontrivial zeros. From the figures we see that the curves of different $\epsilon$ never intersect in a manner similar to the repulsion to level crossing of eigenvalues or the non-crossing of isotherm curves. Since the curves do not cross they do not cross the $\epsilon=0$ curve which is the only way for them to reach the $p$ axis. Another interesting feature is that the $\psi_\epsilon^{i)}(x)$ functions in position space never have a zero. This is a clue that these functions represent ground states of a quantum system which can never have a zero in position space\cite{Susskind:2014qoa}. The ground states in momentum space $\psi_\epsilon^{i)}(x)$ can have a zero but is not required to. It is symmetric in $p$ which is a property of quantum energy states in momentum space. We shall explore more the relation of these functions to ground state wave functions in section 4 where we discuss their relation to supersymmetric quantum  mechanics.
\begin{table}[h]
\centering
\begin{tabular}{|l|l|l|l|l|}
\hline
Function       & Non-trivial zeros & Inverse Fourier Transform \\\hline 
$\tilde \psi_\epsilon(^{(1)}(p) = Ai_4(p+i\epsilon)$&
 Yes &  $\psi_\epsilon^{(1)}(x) =  e^{-\frac{x^4}{256} - \epsilon x} $ \\ \hline
$\tilde \psi_\epsilon ^{(2)}(p) = {K_{ip + \epsilon }}(1) $   & Yes & $\psi_\epsilon ^{(2)}(x) =e^{-\cosh x - \epsilon x}$  \\ \hline
${{\tilde \psi }_\epsilon^{(3)}}(p) = \xi \left( { i p + \frac{1}{2} + \epsilon} \right)$  & Yes & ${\psi^{(3)} _\epsilon}(x) = \Phi (x)e^{-\epsilon x}$   \\ \hline
\end{tabular}
\caption{\label{tab:table-name} Three functions in momentum space with nontrivial zeros: a generalized Airy function, a function of the argument of a K Bessel function and the Riemann xi function.  Ground state wave functions in position space are related to these functions through a Fourier transformation. The introduction of a deformation parameter $\epsilon $ appears to remove the nontrivial zeros in a manner similar to the Riemann hypothesis.}
\end{table}

\begin{figure}[!htb]
\centering
\minipage{0.5\textwidth}
  \includegraphics[width=\linewidth]{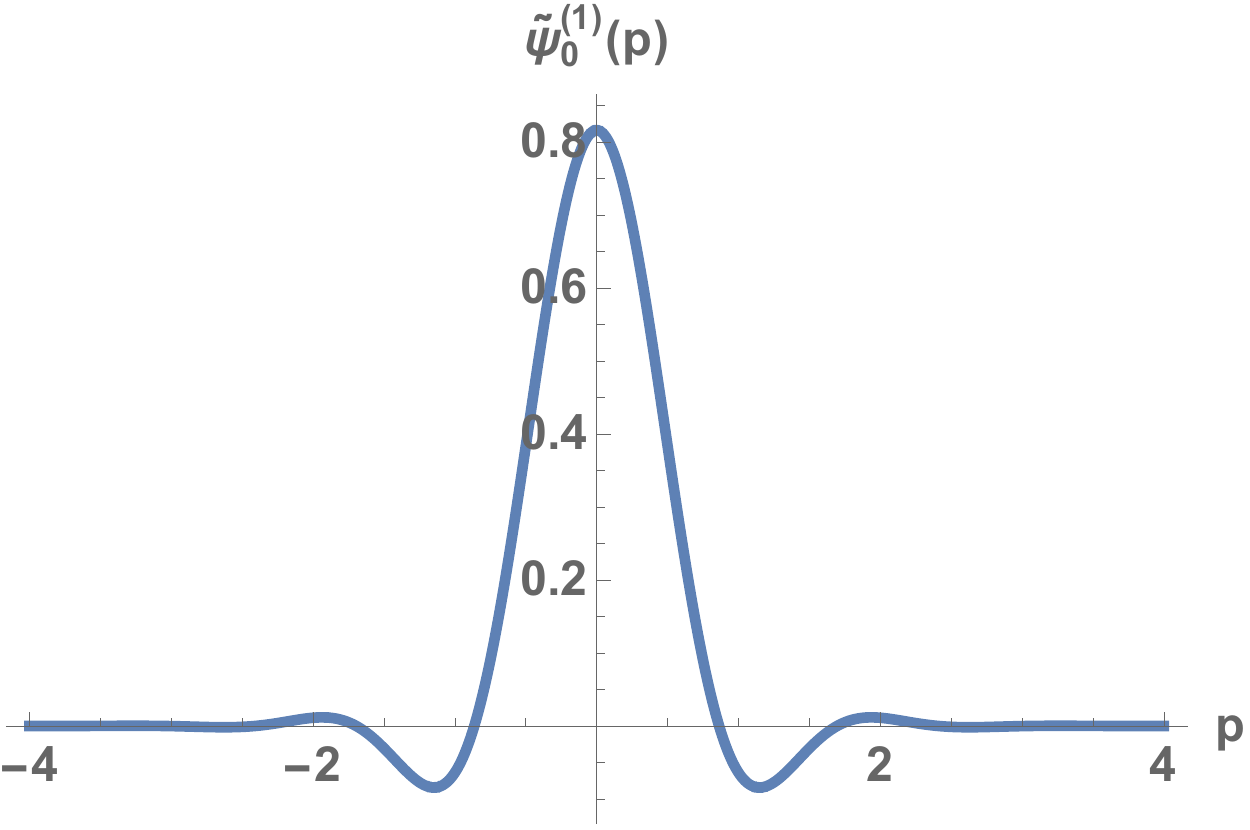}
\endminipage\hfill
\minipage{0.48\textwidth}
  \includegraphics[width=\linewidth]{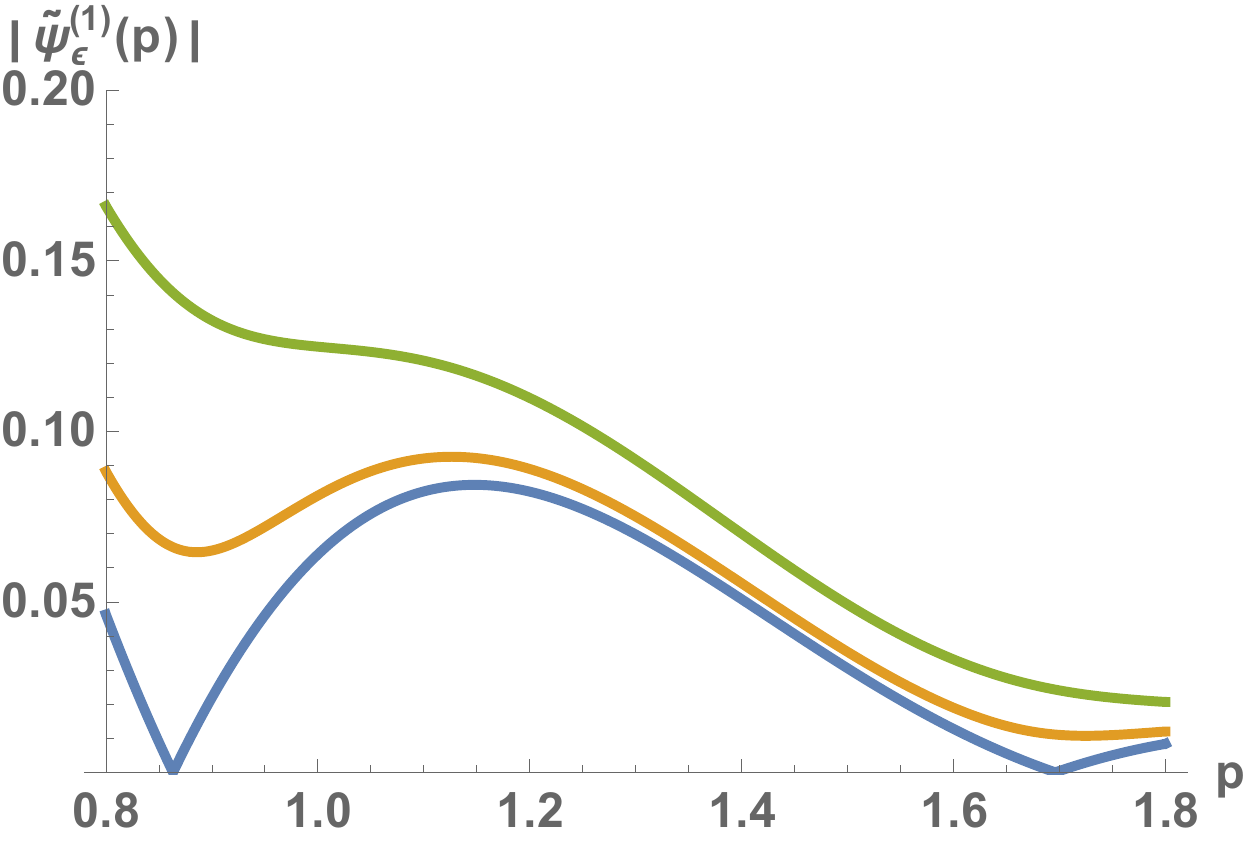}
\endminipage\hfill
\caption{(left) The first function in momentum space with deformation parameter zero (right) Close up view of the magnitude of the first function for deformation parameter $\epsilon = 0$ with zeros at $p_1^{(1)}= .86336603209 $ and $p_2^{(1)}= 1.696081937$ (blue), $\epsilon = .1$ (orange) and $\epsilon = .2$ (green) with no zeros.
}
\end{figure}

\begin{figure}[!htb]
\centering
\minipage{0.5\textwidth}
  \includegraphics[width=\linewidth]{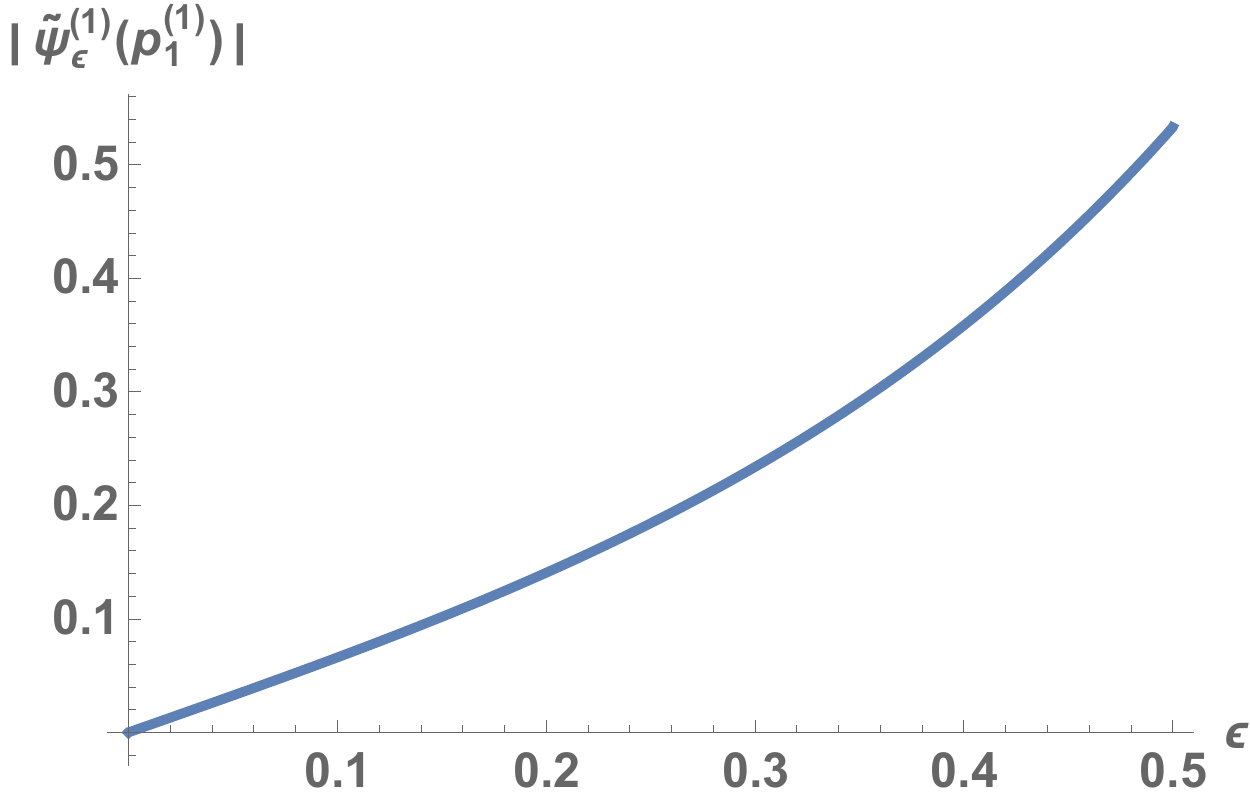}
\endminipage\hfill
\minipage{0.48\textwidth}
  \includegraphics[width=\linewidth]{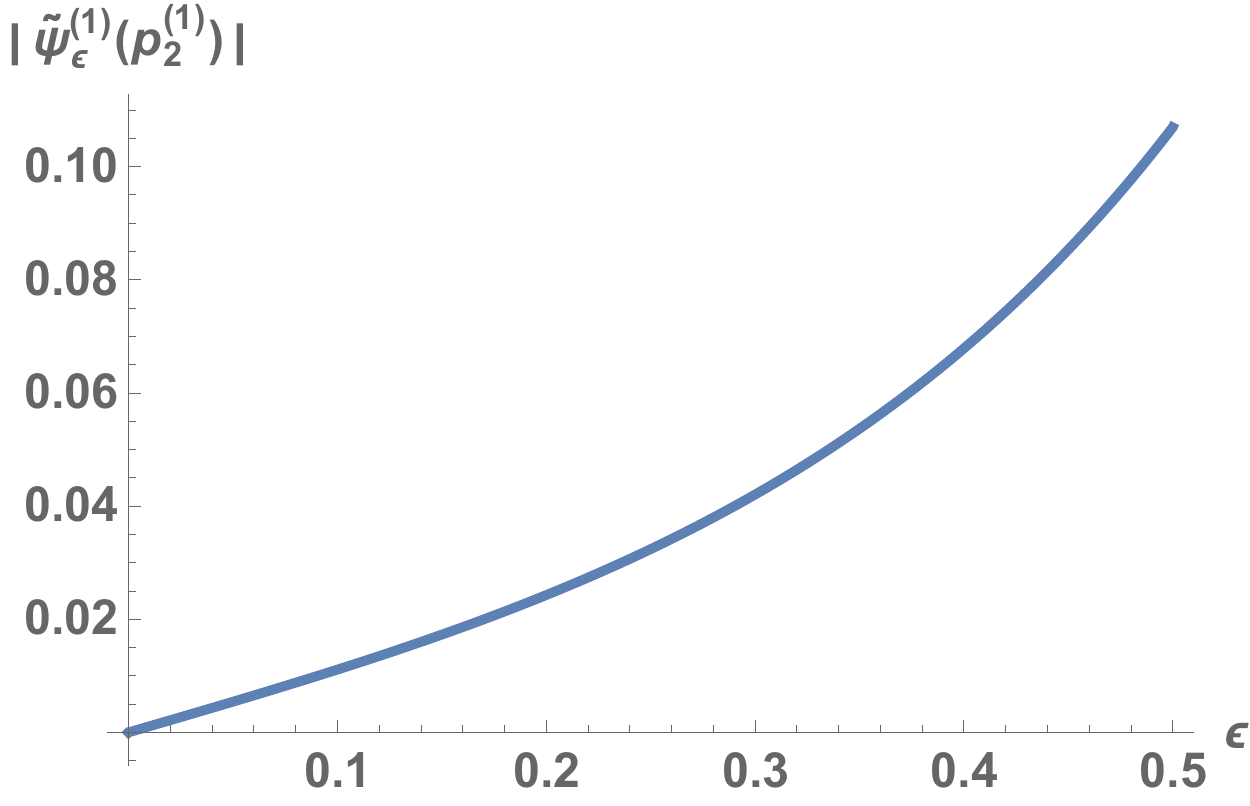}
\endminipage\hfill
\caption{ How the first function moves away from the (left) first zero and (right) second zero as a function of the deformation parameter $\epsilon$. 
}
\end{figure}

\begin{figure}[!htb]
\centering
\minipage{0.5\textwidth}
  \includegraphics[width=\linewidth]{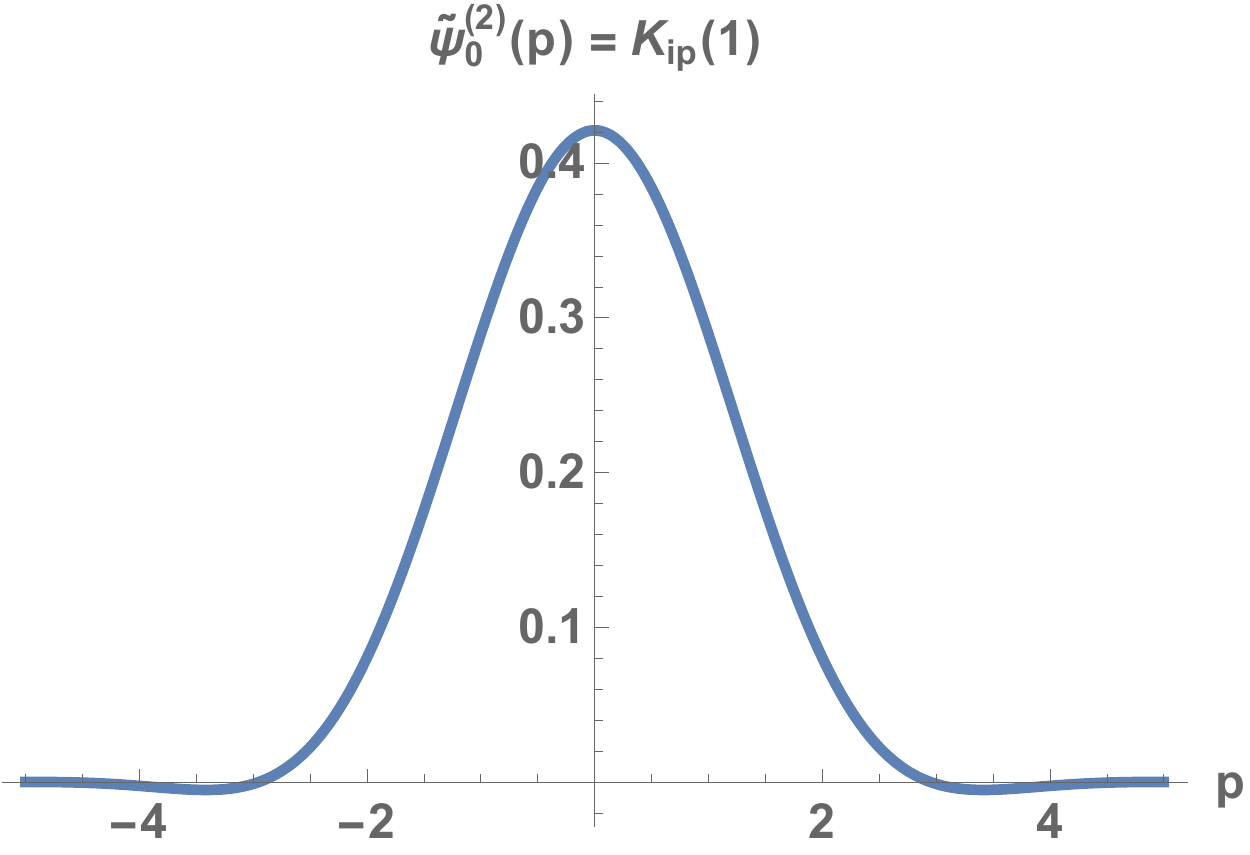}
\endminipage\hfill
\minipage{0.48\textwidth}
  \includegraphics[width=\linewidth]{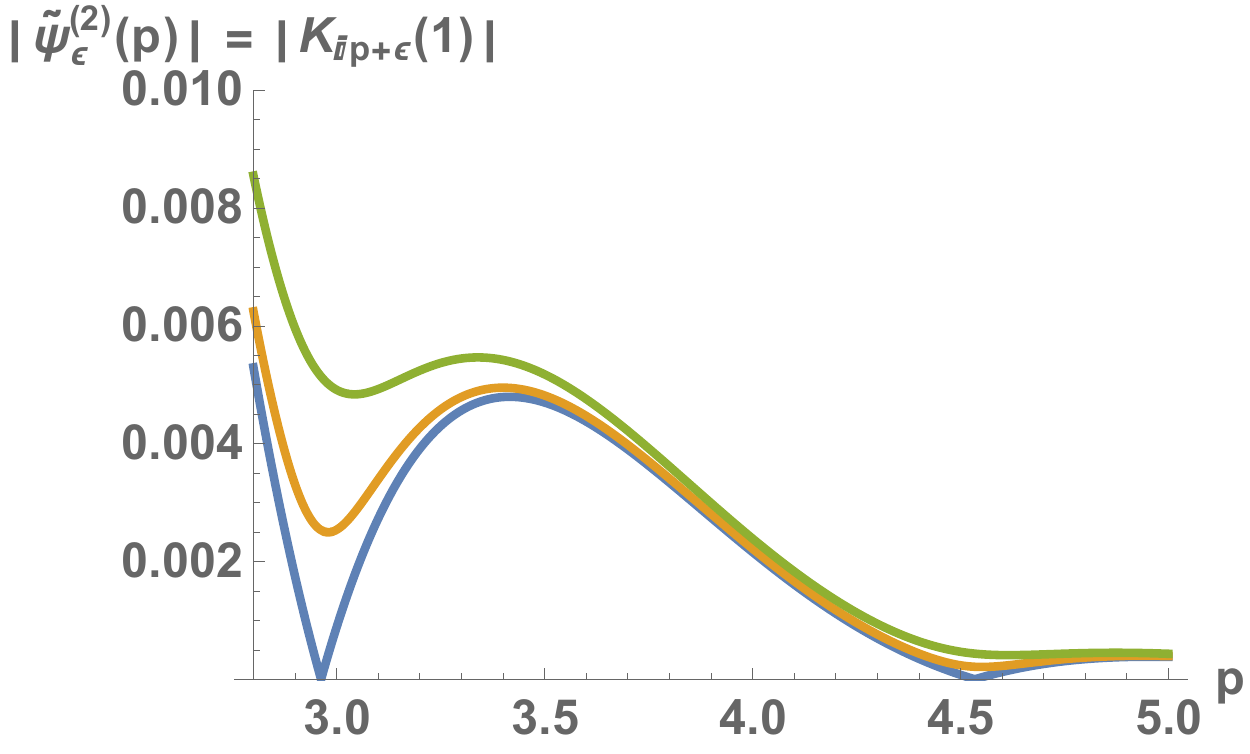}
\endminipage\hfill
\caption{(left) The second function in momentum space with deformation parameter zero (right) Close up view of the magnitude of the second function for deformation parameter $\epsilon = 0$ with zeros at $p_1^{(2)}= 2.962548534571 $ and $p_2^{(2)}= 4.534490718 $ (blue), $\epsilon = .1$ (orange) and $\epsilon = .2$ (green) with no zeros.
}
\end{figure}

\begin{figure}[!htb]
\centering
\minipage{0.5\textwidth}
  \includegraphics[width=\linewidth]{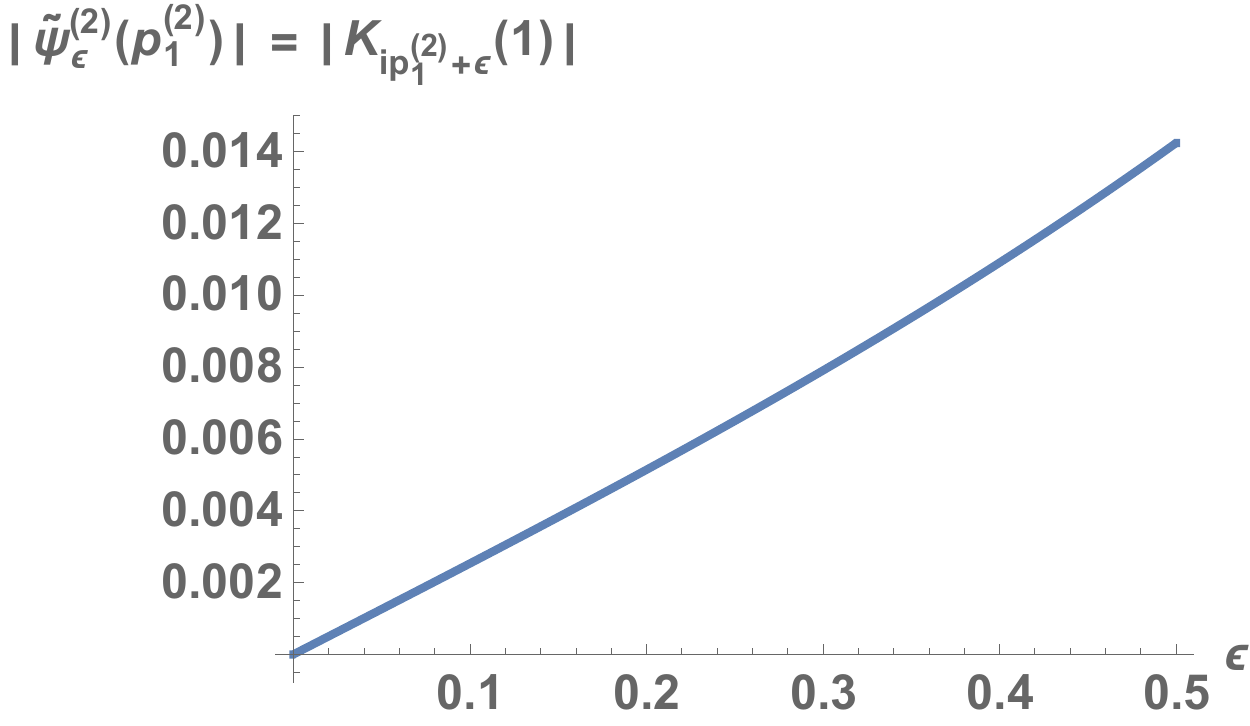}
\endminipage\hfill
\minipage{0.48\textwidth}
  \includegraphics[width=\linewidth]{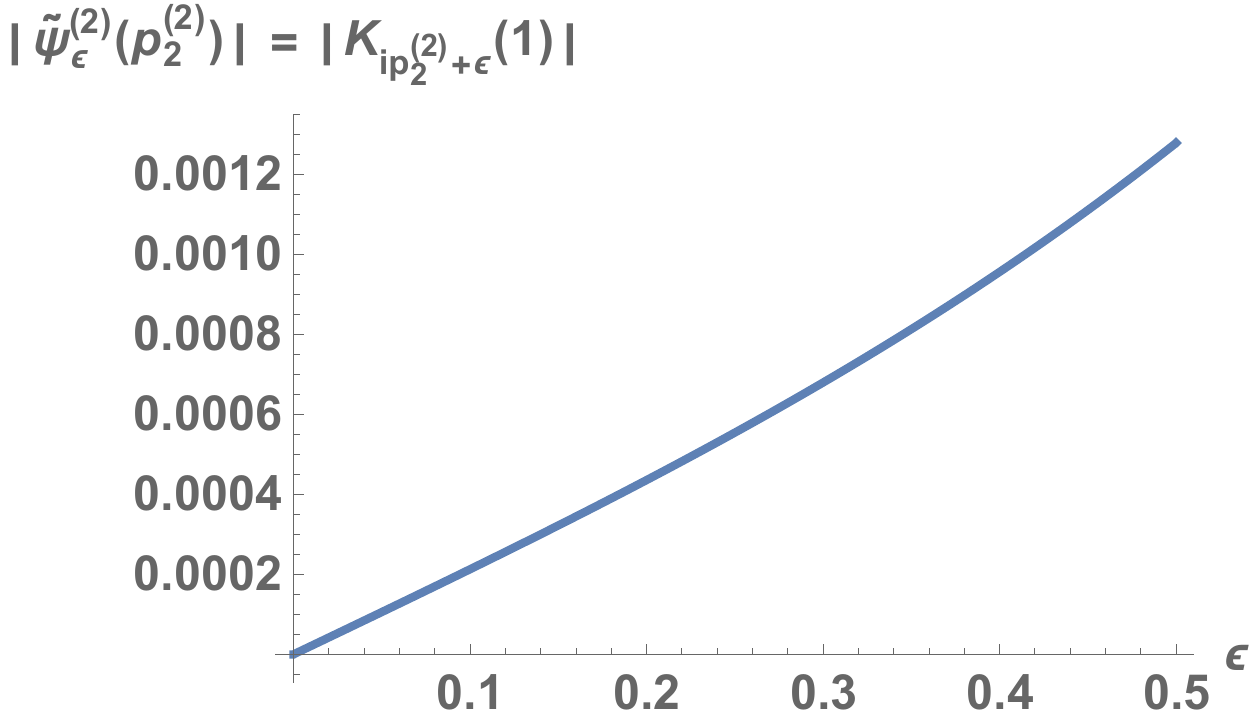}
\endminipage\hfill
\caption{How the second function moves away from the (left) first zero and (right) second zero as a function of the deformation parameter $\epsilon$.
}
\end{figure}

\begin{figure}[!htb]
\centering
\minipage{0.5\textwidth}
  \includegraphics[width=\linewidth]{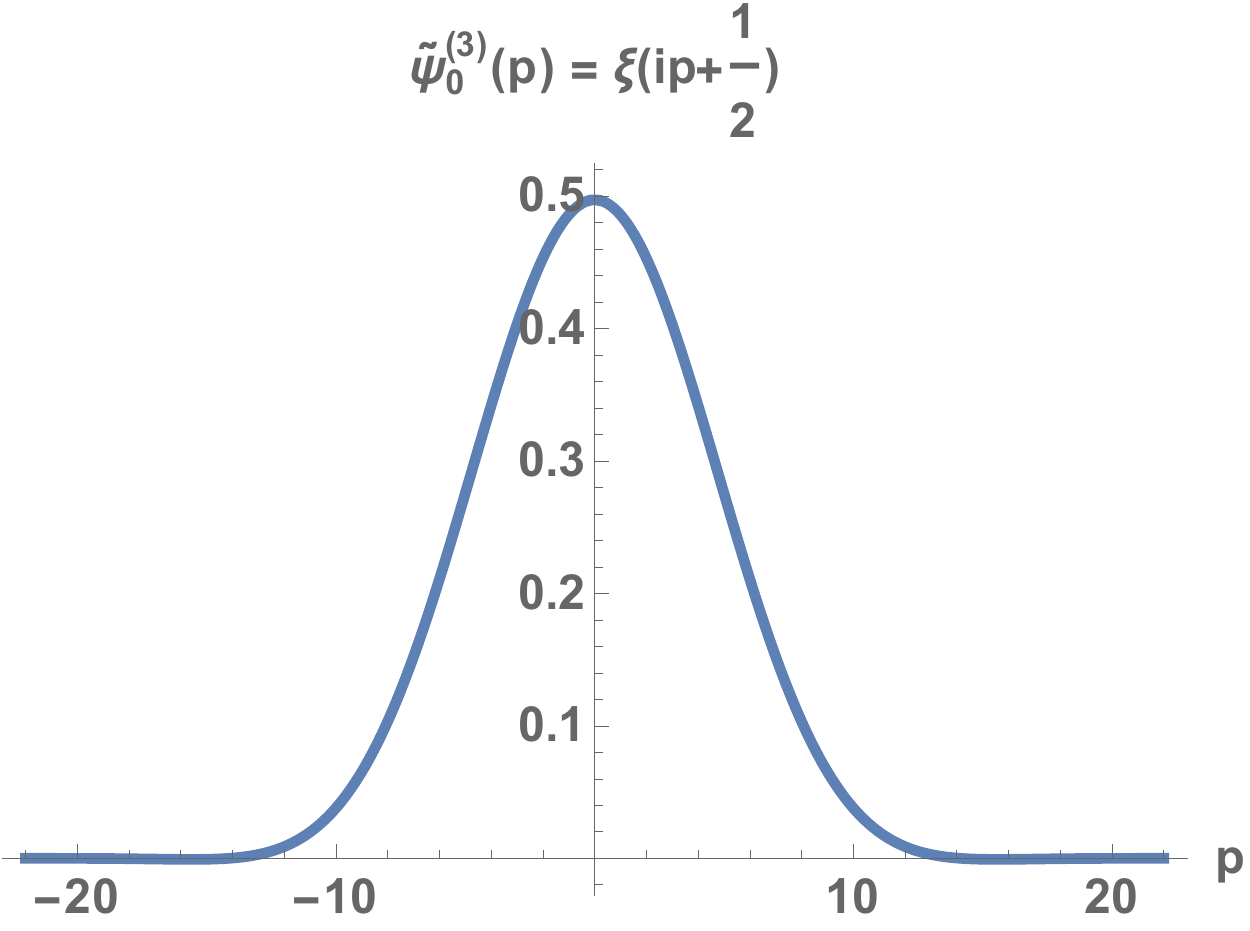}
\endminipage\hfill
\minipage{0.48\textwidth}
  \includegraphics[width=\linewidth]{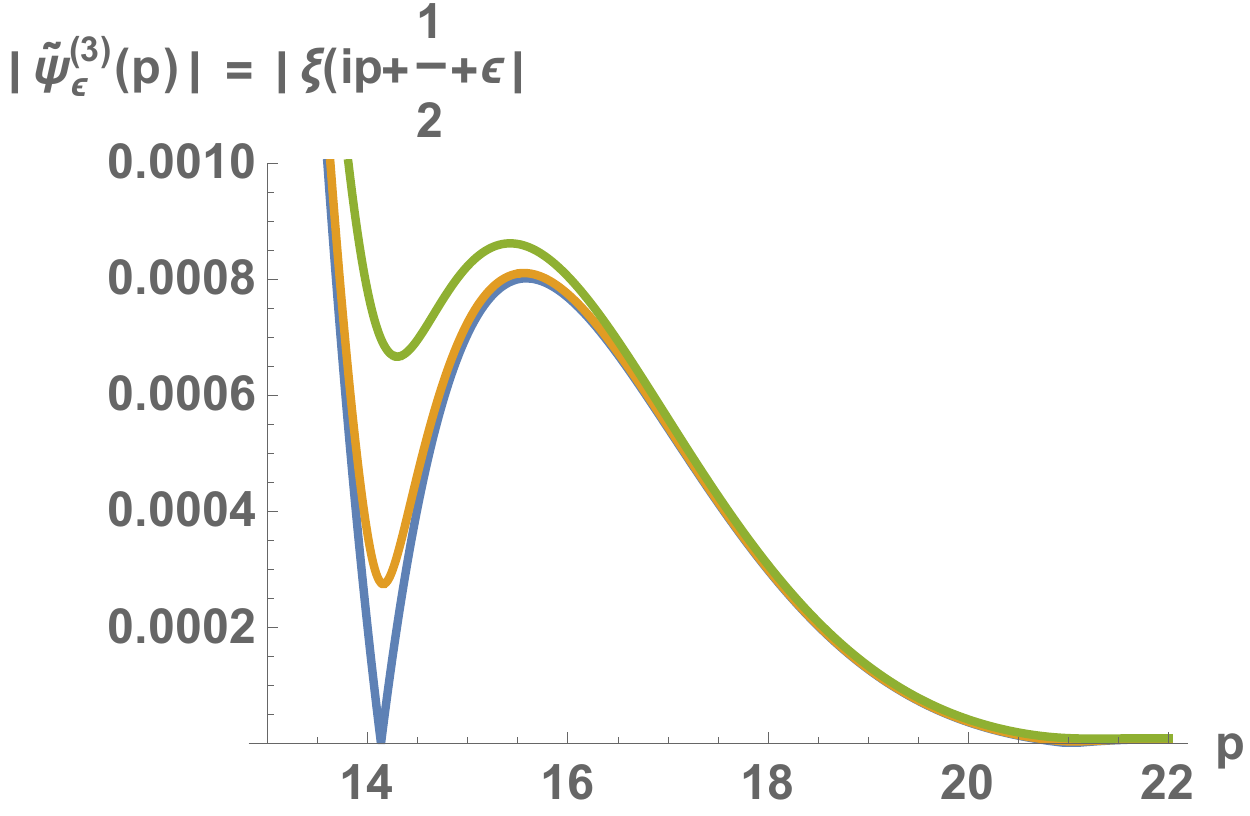}
\endminipage\hfill
\caption{(left) The third function in momentum space with deformation parameter zero (right) Close up view of the magnitude of the second function for deformation parameter $\epsilon = 0$ with zeros at $p_1^{(3)}= 14.134725141734695 $ and $p_2^{(3)}= 21.022039638771556 $ (blue), $\epsilon = .2$ (orange) and $\epsilon = .5$ (green) with no zeros.
}
\end{figure}

\begin{figure}[!htb]
\centering
\minipage{0.5\textwidth}
  \includegraphics[width=\linewidth]{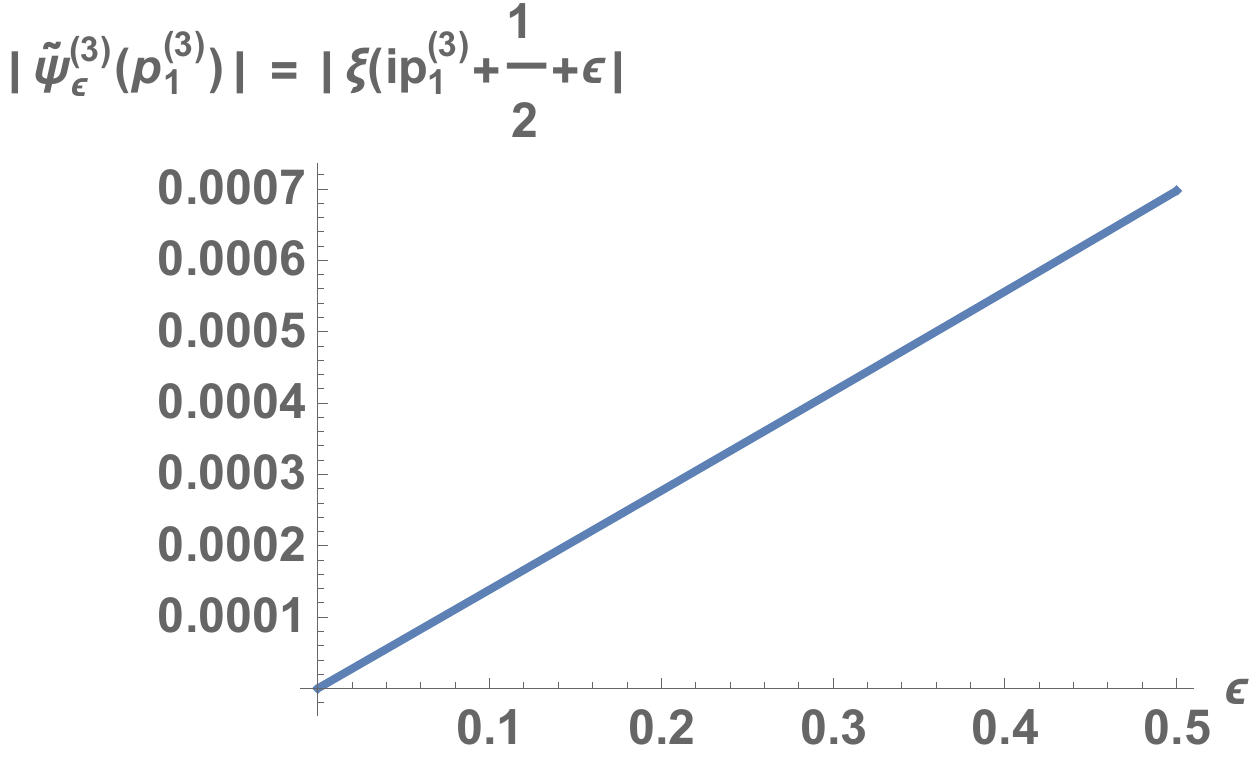}
\endminipage\hfill
\minipage{0.48\textwidth}
  \includegraphics[width=\linewidth]{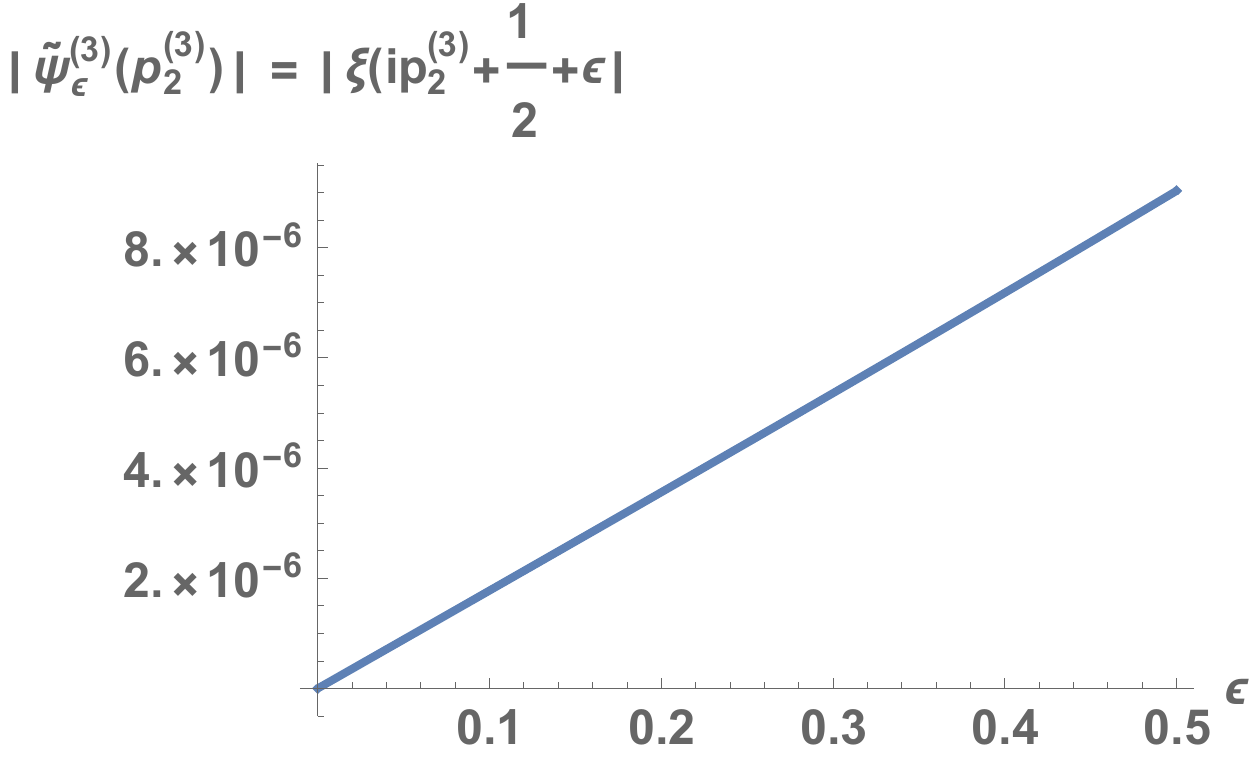}
\endminipage\hfill
\caption{ How the third function moves away from the (left) first zero and (right) second zero as a function of the deformation parameter $\epsilon$.
}
\end{figure}

\section{Quantum Fourier transform on a quantum computer}

The Quantum Fourier Transform (QFT) algorithm is based on a quantum computers innate ability to represent the operation of Unitary matrices on quantum states. On a quantum computer both the Unitary matrix and the states are represented in terms of qubits  which are acted on but quantum gates and represented in a quantum circuit. The Unitary matrix that the QFT algorithm represents in the  Sylvester matrix $F$. In terms of quantum systems this matrix allows one to transform from a discrete version of states in position space to a discrete version of states in momentum space. In the position basis the position matrix $Q_{pos}$ is given by the diagonal matrix:
\begin{equation}
{\left( {{Q_{pos}}} \right)_{j,k}} = \sqrt {\frac{{2\pi }}{{4N}}} (2j - (N + 1)){\delta _{j,k}}
\end{equation}
and the momentum matrix $P_{pos}$ obtained from this through the operation of the Sylvester matrix by \cite{Zhang:2004ct}:
\begin{equation}{P_{pos}} = {F^\dag }{Q_{pos}}F\end{equation}
where 
\begin{equation}{F_{j,k}} = \frac{1}{{\sqrt N }}{e^{\frac{{2\pi i}}{{4N}}(2j - (N + 1))(2k - (N + 1))}}\end{equation}
In the same way quantum states in the position space are related to the states in momentum basis through the operation of $F$ which implements the QFT. A quantum circuit we used to apply the QFT to our functions is shown in figure 7. We used six qubits in our computations hence the six horizontal lines in the diagram. The six qubits represented 64 data points for the initial and transformed state. Also note the high depth nature of the QFT circuit as there are many gates and the circuit stretches out far in the horizontal direction. Nevertheless we were able to see the proper structure of the three functions as well as position of the first zeros. More details of the functions can emerge as more qubits and longer coherence times become feasible. The Quantum Fourier Transform is potentially exponentially faster on a fault tolerant quantum computer than the classical Fast Fourier Transform algorithm. It is one of the core quantum computing algorithms on which many important algorithms, for example Shor's factoring algorithm, is based. The procedure we  use for the QFT algorithm is to go between the position and momentum ground states associated with the three functions above and investigate the ability in determining the zeros of the functions. The QFT works by representing the Sylvester matrix as a Unitary matrix, expanding the Unitary matrix systematically in terms of gates and using these gates to form a quantum circuit. This quantum circuit is evaluated with discrete data in terms of the quantum ground state in position space with the output yielding the quantum state in momentum space, also represented in terms of discrete data and qubits. The results of the QFT are then analyzed to see if they are accurate enough to determine the zeros of the functions and how they move away from zero as one turns on the deformation parameter $\epsilon$. Our QFT simulations were done using the IBM QISkit state vector simulator. The same quantum computations can be done on near term quantum computers such as those using superconducting qubits but there is noise and errors present in the gates which cause errors in the QFT \cite{SierraIBM}. Recently however a simple type of quantum error correction has been implemented by Google \cite{GoogleQuantumAI:2022fyn} which may improve the accuracy running on quantum hardware in the near future. Our results are recorded in figures 8-19. We were able to use the QFT method to determine the zeros of the three functions although the results were difficult to pick out because of the small value of the functions near the zeros. In particular for the third function it was necessary to take the log of the absolute value of the data points to see the existence and location of the zeros ( see figure 18-19).

\begin{figure}
\centering
  \includegraphics[width = .5 \linewidth]{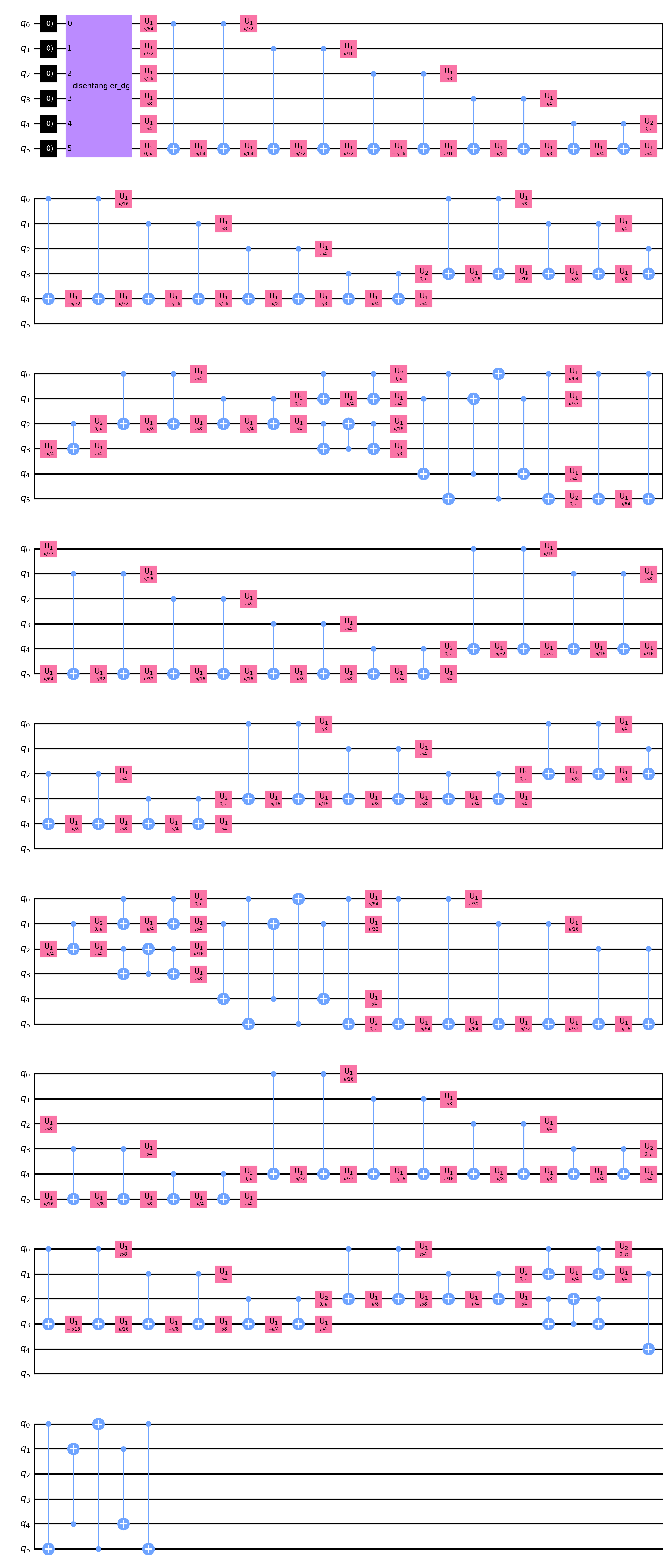}
  \caption{Quantum circuit used to implement the Quantum Fourier Transform (QFT) with six qubits using the IBM QISKit state vector simulator. Because of the length of the cirtcuit the diagram was broken in nine pieces for ease of viewing.}
  \label{fig:Radion Potential}
\end{figure}

\begin{figure}[!htb]
\centering
\minipage{0.5\textwidth}
  \includegraphics[width=\linewidth]{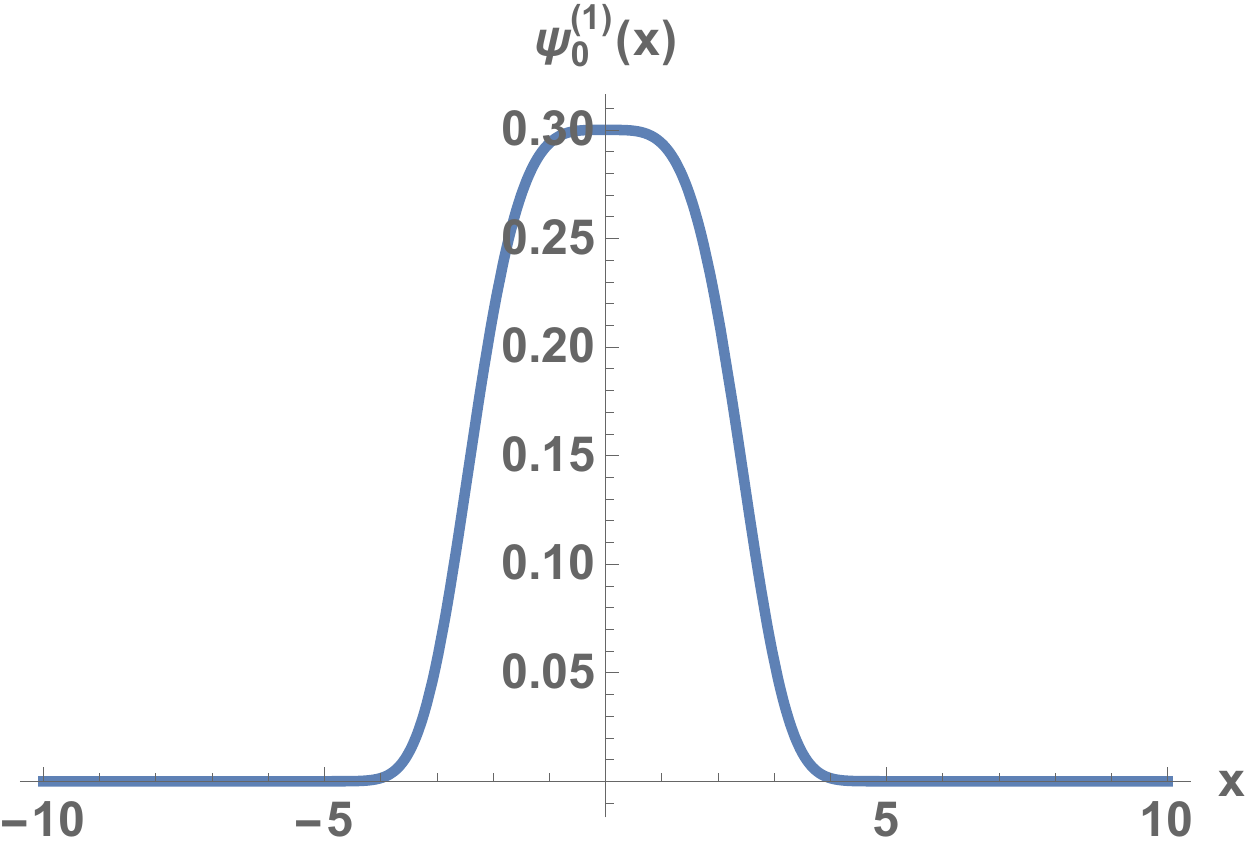}
\endminipage\hfill
\minipage{0.48\textwidth}
  \includegraphics[width=\linewidth]{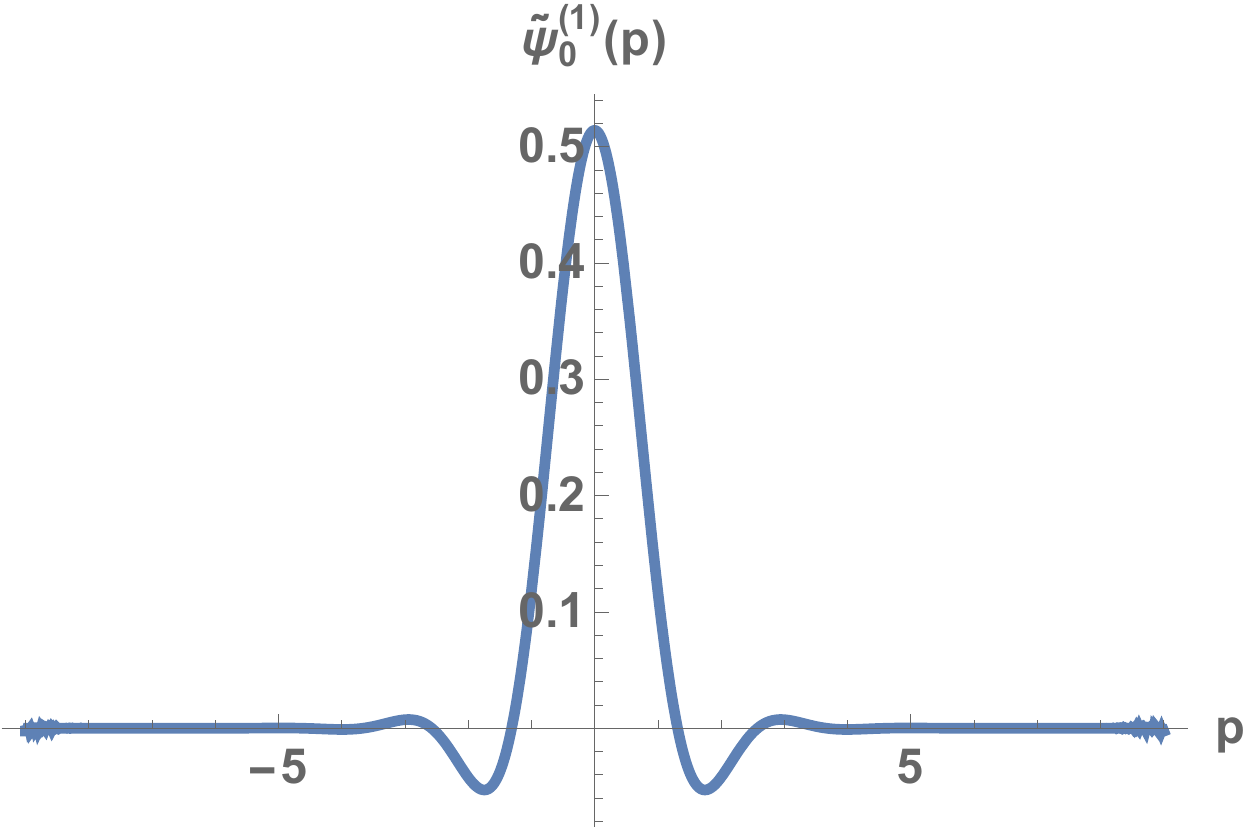}
\endminipage\hfill
\caption{Ground state wave function (left) in position space and (right) in momentum space after Fourier transformation with deformation parameter $\epsilon = 0$. 
}
\end{figure}

\begin{figure}[!htb]
\centering
\minipage{0.5\textwidth}
  \includegraphics[width=\linewidth]{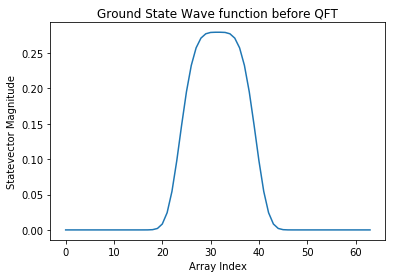}
\endminipage\hfill
\minipage{0.48\textwidth}
  \includegraphics[width=\linewidth]{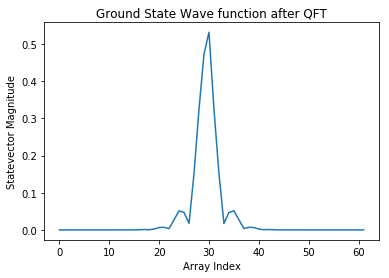}
\endminipage\hfill
\caption{Magnitude of the ground state wave function (left) in position space and (right) in momentum space after Quantum Fourier transformation with deformation parameter $\epsilon = 0$. }
\end{figure}

\begin{figure}[!htb]
\centering
\minipage{0.5\textwidth}
  \includegraphics[width=\linewidth]{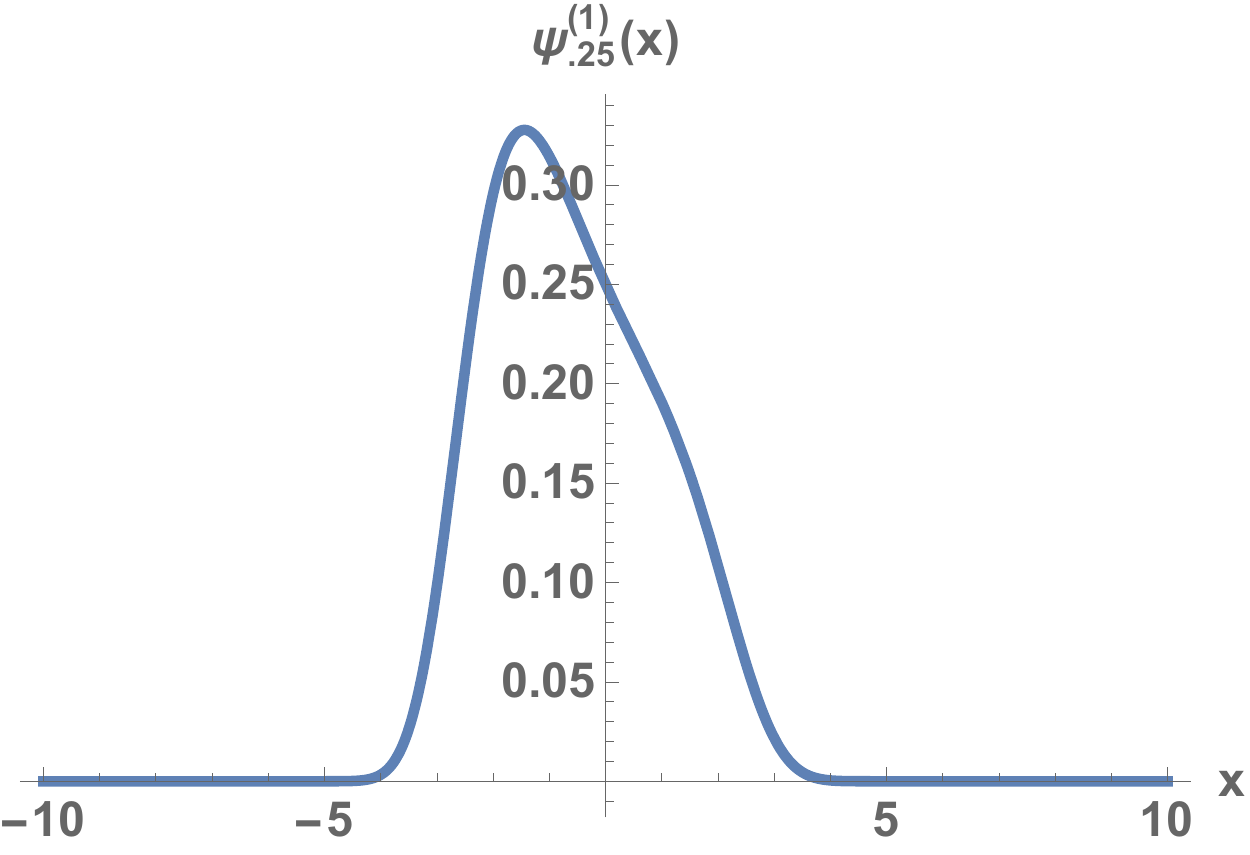}
\endminipage\hfill
\minipage{0.48\textwidth}
  \includegraphics[width=\linewidth]{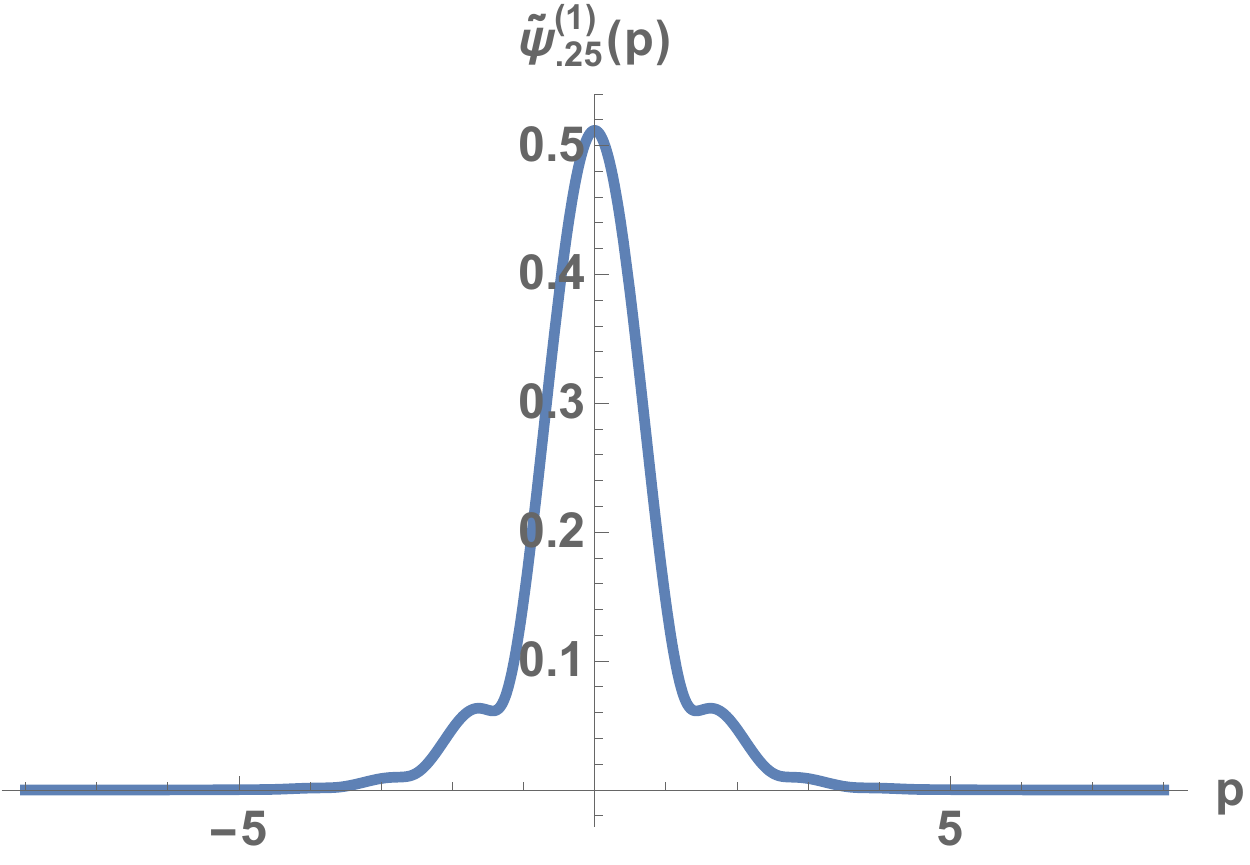}
\endminipage\hfill
\caption{Magnitude of the first ground state wave function (left) in position space and (right) in momentum space after Fourier transformation with deformation parameter $\epsilon = .25$. 
}
\end{figure}

\begin{figure}[!htb]
\centering
\minipage{0.5\textwidth}
  \includegraphics[width=\linewidth]{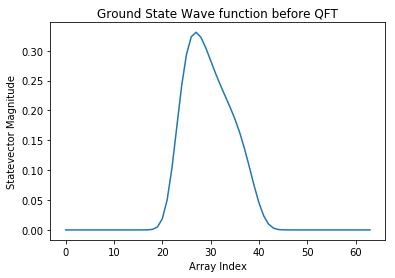}
\endminipage\hfill
\minipage{0.48\textwidth}
  \includegraphics[width=\linewidth]{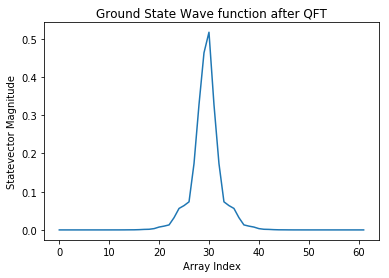}
\endminipage\hfill
\caption{Magnitude of the first ground state wave function (left) in position space and (right) in momentum space after Quantum Fourier transformation with deformation parameter $\epsilon = .25$. 
}
\end{figure}

\begin{figure}[!htb]
\centering
\minipage{0.5\textwidth}
  \includegraphics[width=\linewidth]{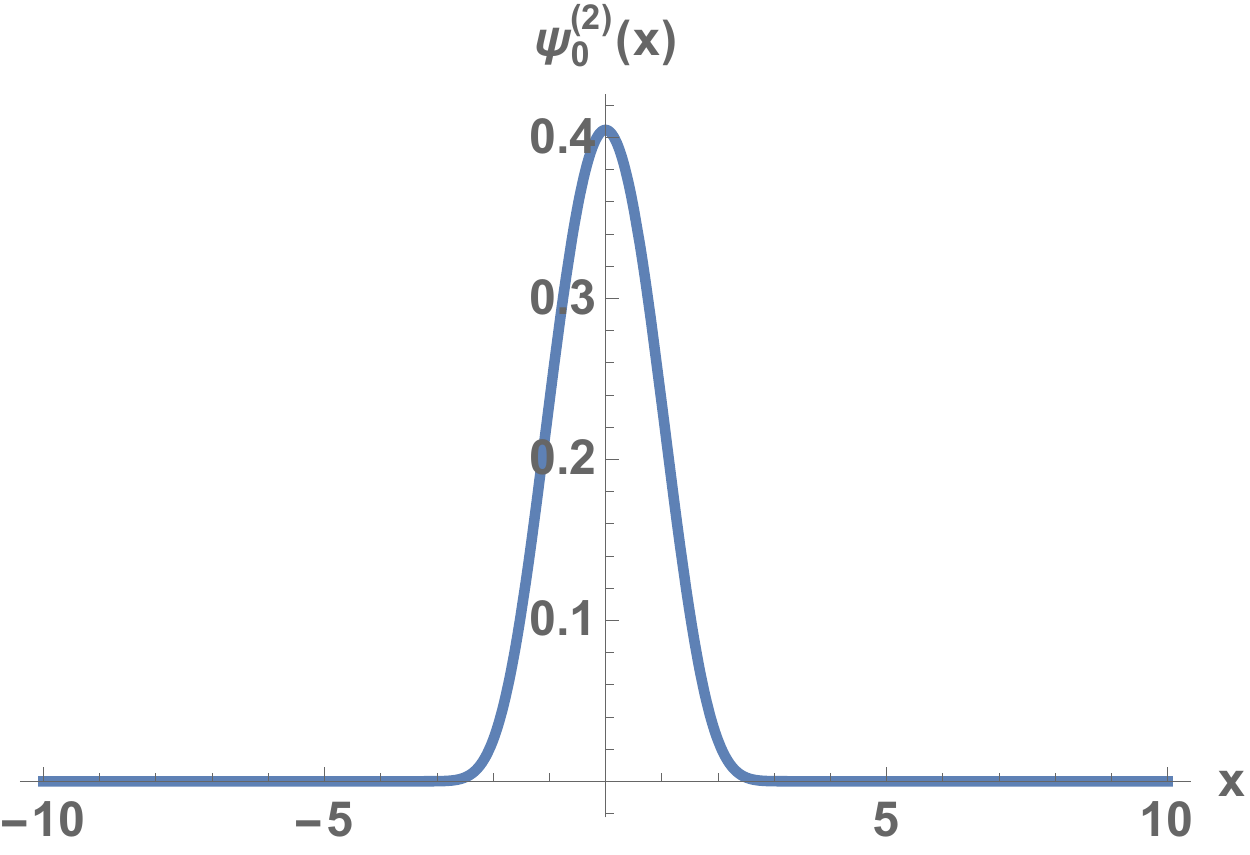}
\endminipage\hfill
\minipage{0.48\textwidth}
  \includegraphics[width=\linewidth]{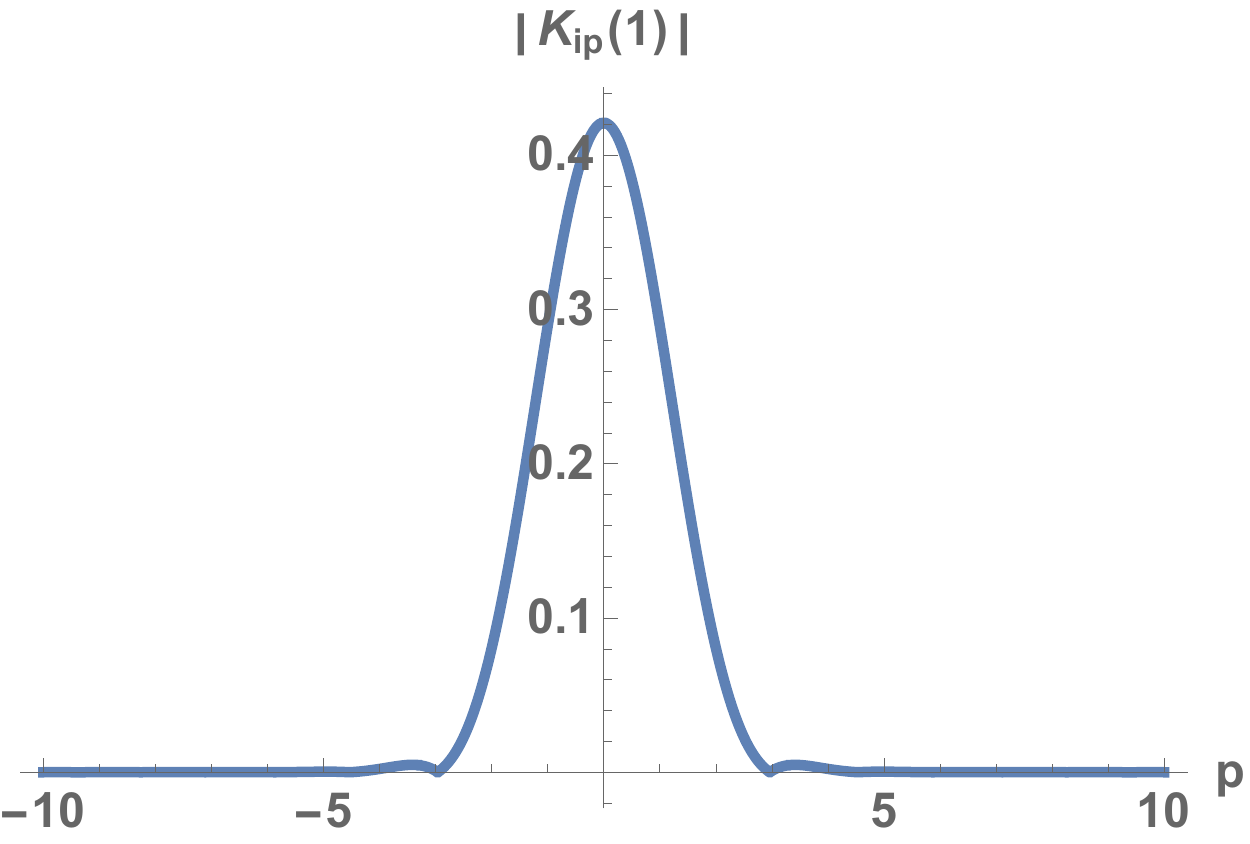}
\endminipage\hfill
\caption{Magnitude of the second ground state wave function (left) in position space and (right) in momentum space after Fourier transformation with deformation parameter $\epsilon = 0$. 
}
\end{figure}

\begin{figure}[!htb]
\centering
\minipage{0.5\textwidth}
  \includegraphics[width=\linewidth]{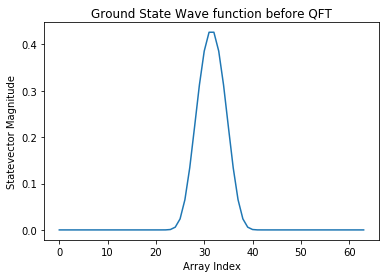}
\endminipage\hfill
\minipage{0.48\textwidth}
  \includegraphics[width=\linewidth]{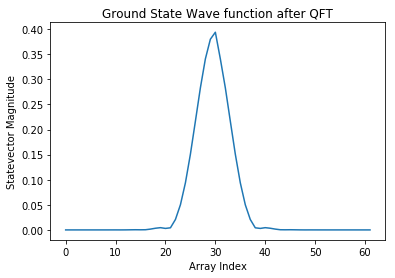}
\endminipage\hfill
\caption{Magnitude of the second ground state wave function (left) in position space and (right) in momentum space after Quantum Fourier transformation with deformation parameter $\epsilon = 0$. 
}
\end{figure}

\begin{figure}[!htb]
\centering
\minipage{0.5\textwidth}
  \includegraphics[width=\linewidth]{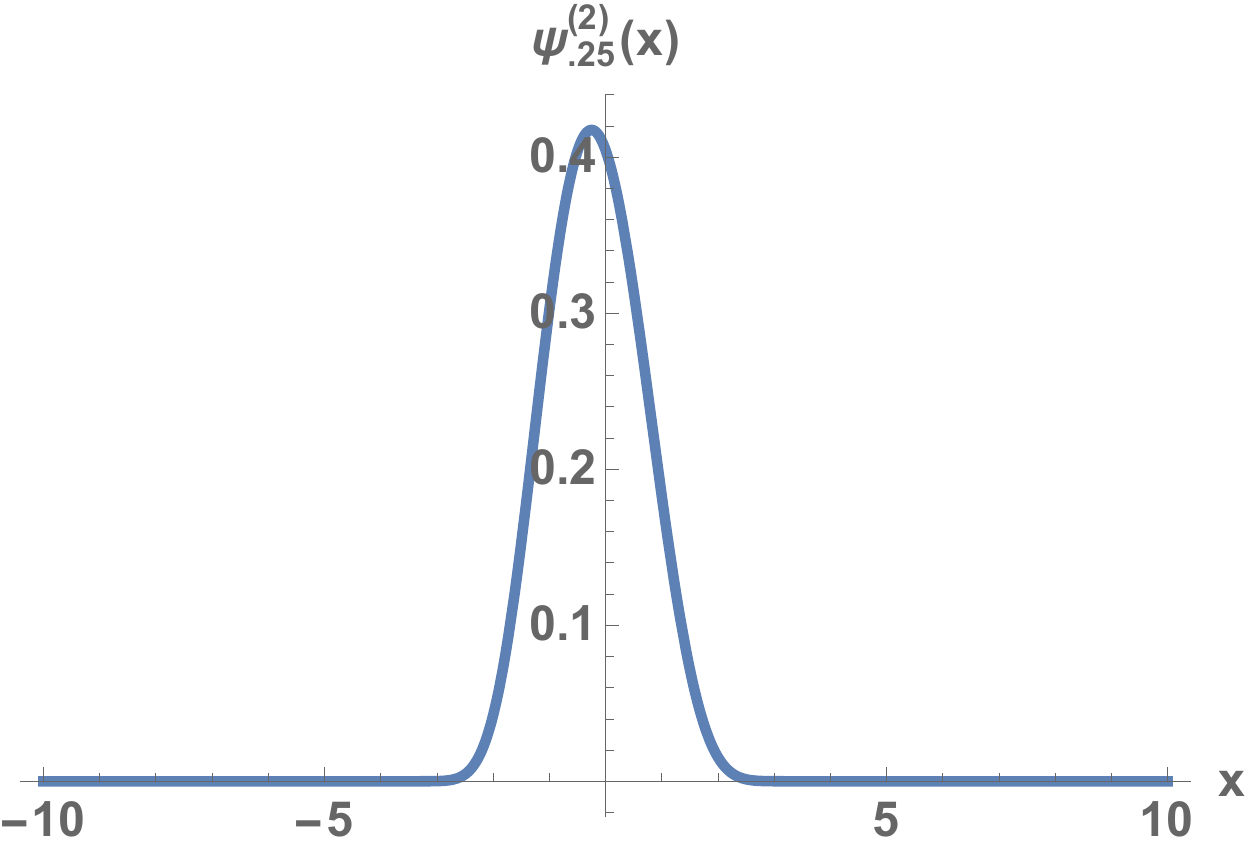}
\endminipage\hfill
\minipage{0.48\textwidth}
  \includegraphics[width=\linewidth]{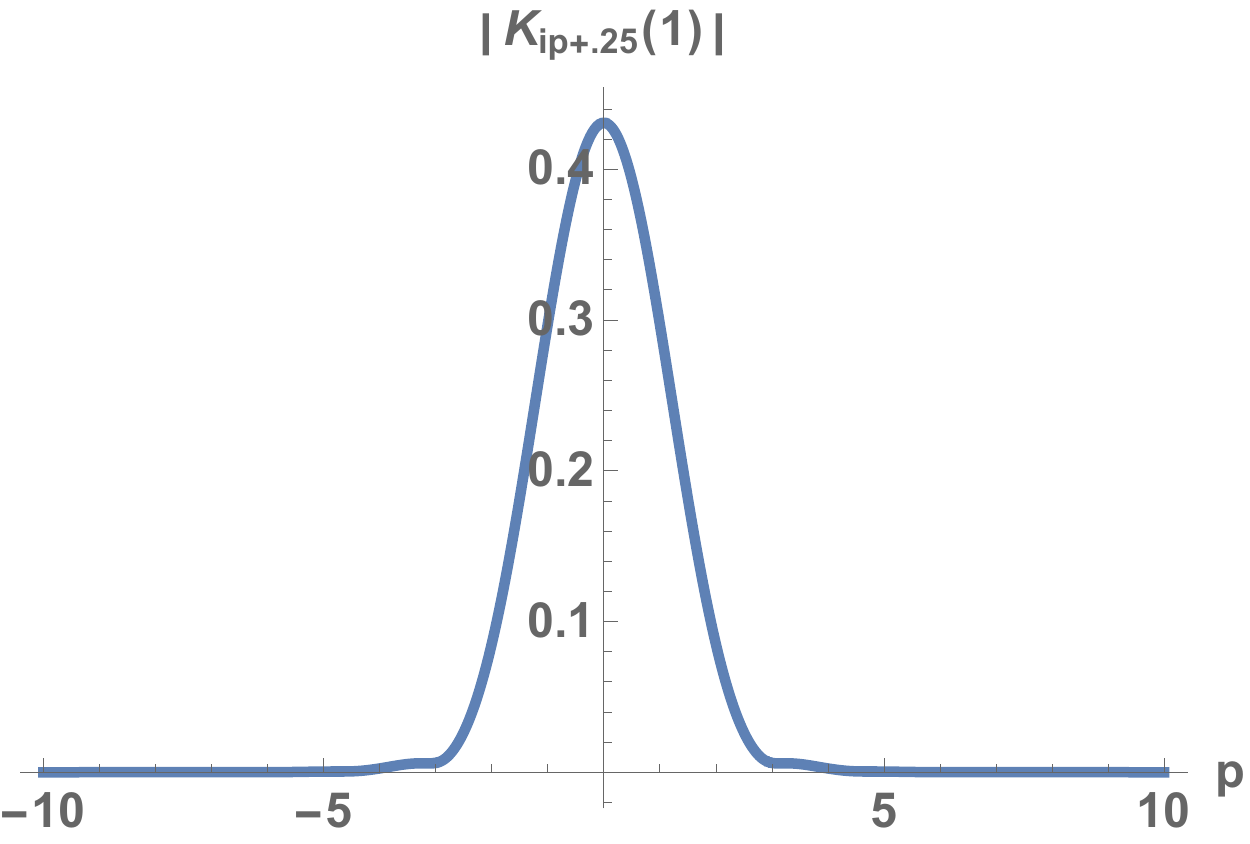}
\endminipage\hfill
\caption{Magnitude of the second ground state wave function (left) in position space and (right) in momentum space after Fourier transformation with deformation parameter $\epsilon = .25$. 
}
\end{figure}

\begin{figure}[!htb]
\centering
\minipage{0.5\textwidth}
  \includegraphics[width=\linewidth]{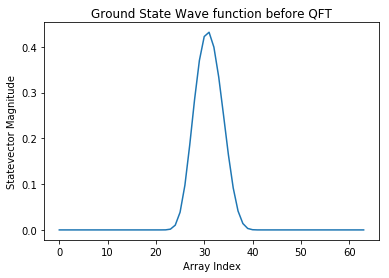}
\endminipage\hfill
\minipage{0.48\textwidth}
  \includegraphics[width=\linewidth]{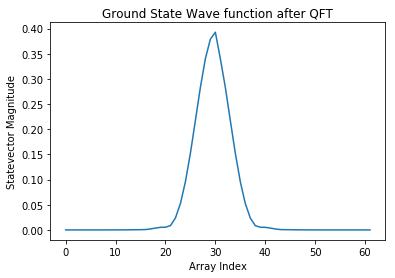}
\endminipage\hfill
\caption{Magnitude of the second ground state wave function (left) in position space and (right) in momentum space after Quantum Fourier transformation with deformation parameter $\epsilon = .25$. 
}
\end{figure}

\begin{figure}[!htb]
\centering
\minipage{0.5\textwidth}
  \includegraphics[width=\linewidth]{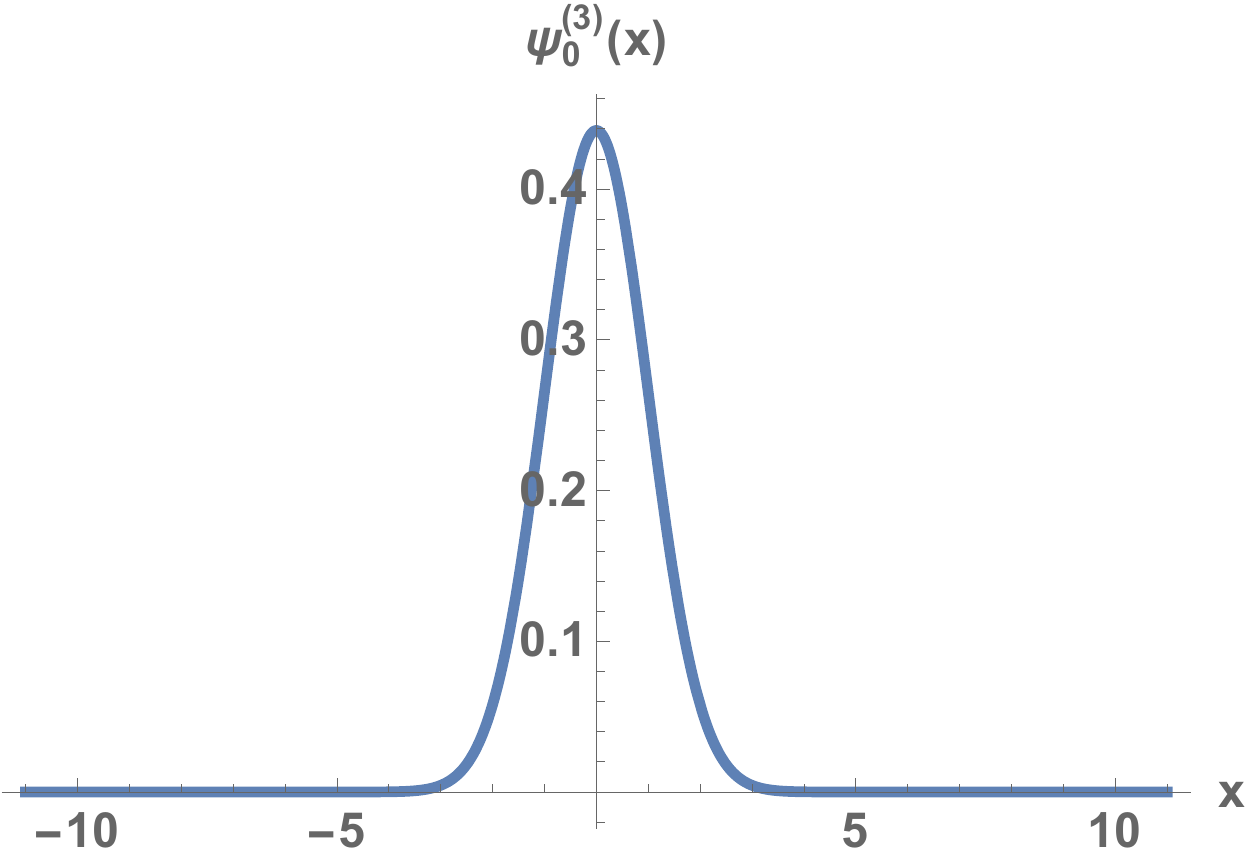}
\endminipage\hfill
\minipage{0.48\textwidth}
  \includegraphics[width=\linewidth]{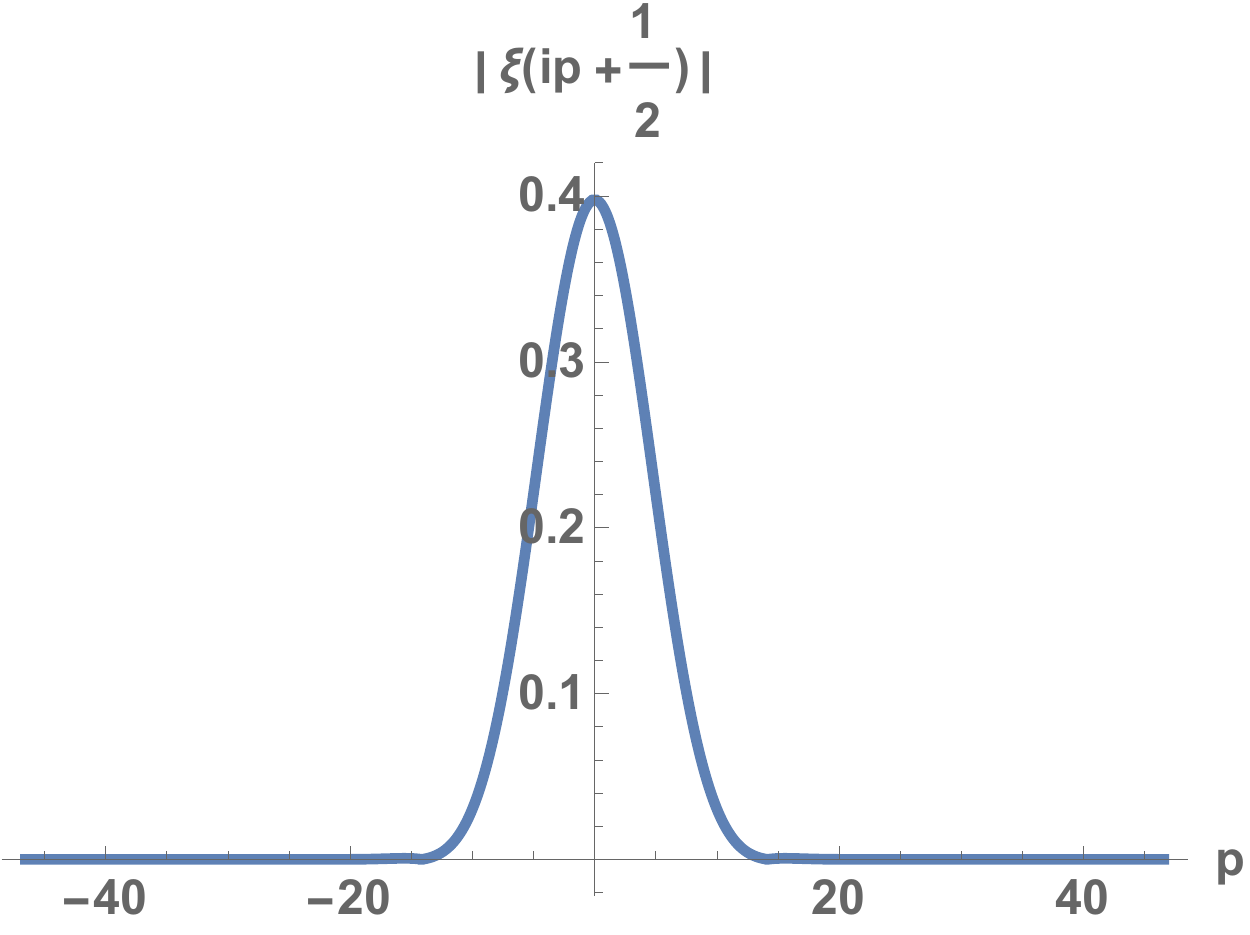}
\endminipage\hfill
\caption{Magnitude of the third ground state wave function (left) in position space and (right) in momentum space after Fourier transformation with deformation parameter $\epsilon = 0$.
}
\end{figure}

\begin{figure}[!htb]
\centering
\minipage{0.48\textwidth}
  \includegraphics[width=\linewidth]{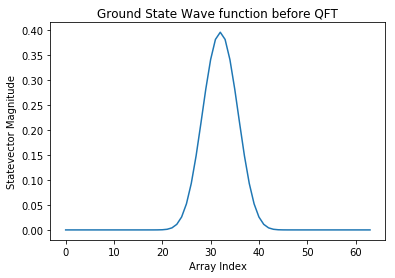}
\endminipage\hfill
\minipage{0.48\textwidth}
  \includegraphics[width=\linewidth]{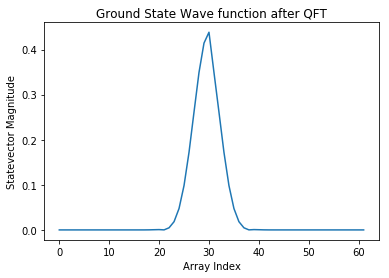}
\endminipage\hfill
\caption{Magnitude of the third ground state wave function (left) in position space and (right) in momentum space after Quantum Fourier transformation with deformation parameter $\epsilon = 0$.
}
\end{figure}

\begin{figure}[!htb]
\centering
\minipage{0.48\textwidth}
  \includegraphics[width=\linewidth]{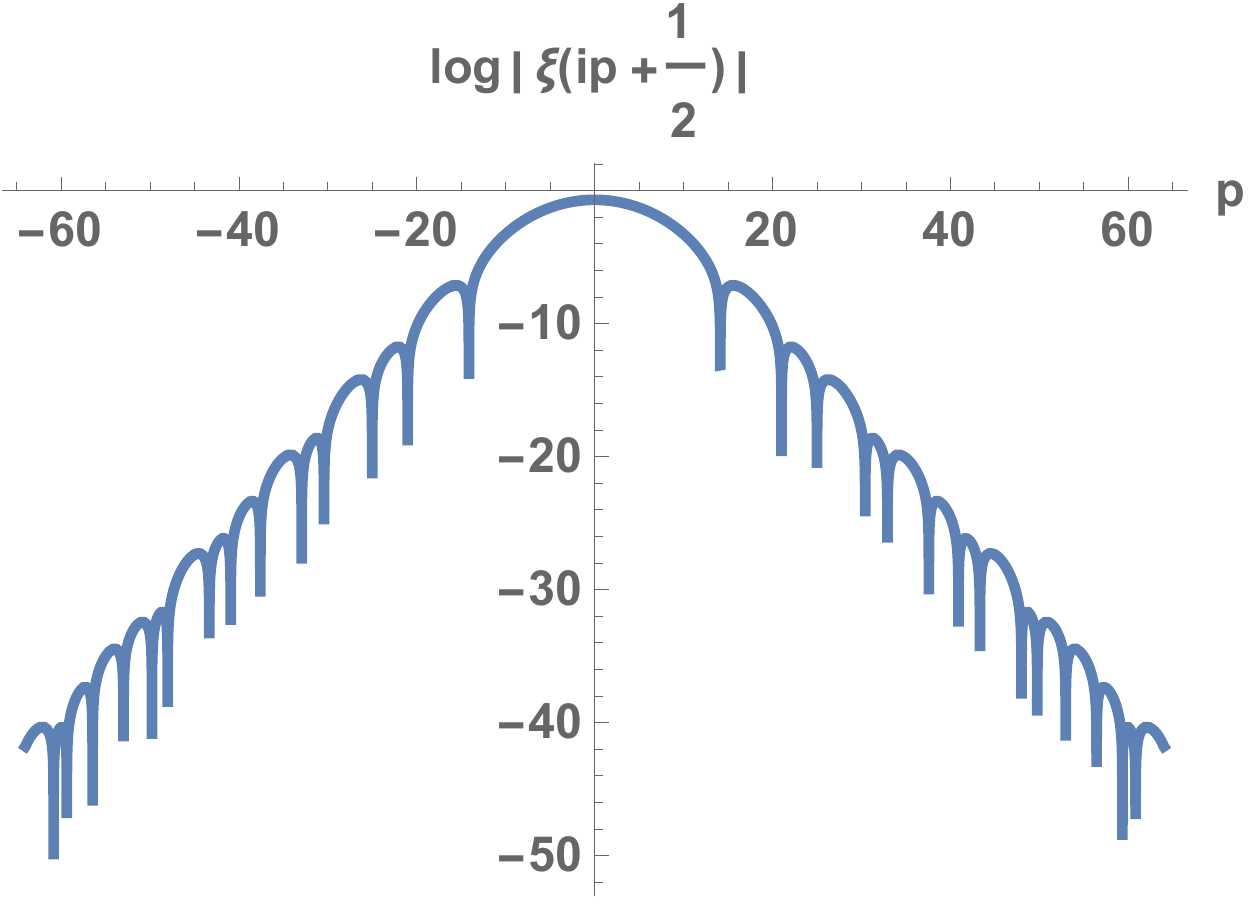}
\endminipage\hfill
\minipage{0.48\textwidth}
\vspace{.8cm}
  \includegraphics[width=\linewidth]{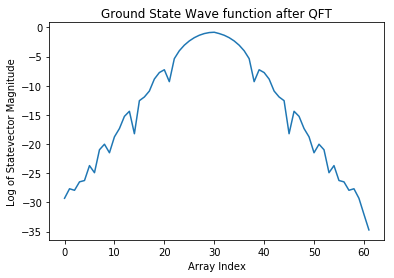}
\endminipage\hfill
\caption{Log of the magnitude of the third ground state wave function in momentum space (left) using the exact expression and (right)using the Quantum Fourier transformation with deformation parameter $\epsilon = 0$.
}
\end{figure}

\begin{figure}[!htb]
\centering
\minipage{0.48\textwidth}
  \includegraphics[width=\linewidth]{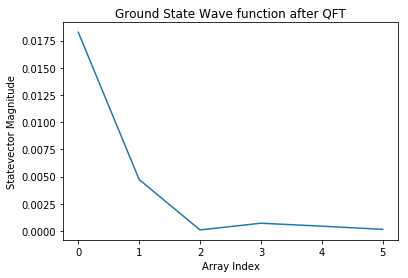}
\endminipage\hfill
\minipage{0.48\textwidth}
  \includegraphics[width=\linewidth]{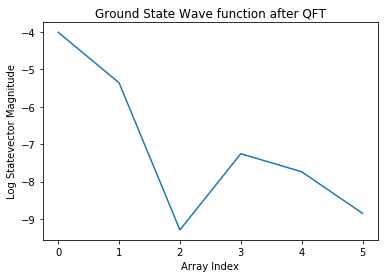}
\endminipage\hfill
\caption{Zoom in on the (left) the magnitude and (right) log of the magnitude of the third ground state wave function in momentum space using the Quantum Fourier transformation with deformation parameter $\epsilon = 0$.
}
\end{figure}

\section{Relation to supersymmetric quantum mechanics}

Recently there has been several investigations of the interesting relation of supersymmetric quantum mechanics to the Riemann hypothesis \cite{Das:2018dit}
\cite{Kalauni:2022cdj}
\cite{Garcia-Munoz:2023kju}
\cite{McGuigan:2020lsp}.  All three of the functions discussed above can be realized as ground state wave functions in supersymmetric quantum mechanics (SUSY-QM). In supersymmetric quantum mechanics the superpotential plays a key role \cite{Gangopadhyaya:2017wpf}. There are two partner potentials and Hamiltonians (plus and minus) which are determined from the superpotential. The minus partner potential determines the the ground state wave function.  The minus partner potential $V_(x)$ is determined from the superpotential $W(x)$ by:
\begin{equation}{V_ - }(x) = {W^2}(x) - W'(x)\end{equation}
The  ground state wave function is determined exactly in SUSYQM also from the superptential through
\begin{equation}\psi (x) = {e^{ - \int_0^x {W(x')dx'} }}\end{equation}
The reason for this simple expression is that the Hamiltonian in SUSY-QM is expressed as the product of two operators which are first order in momentum and allows the ground state to be solved for in terms of the superpotential.

For the first function $\psi_\epsilon^{(1)}(x)$ the superpotential is 
\begin{equation}{W^{(1)}}(x) = g{x^3} + \epsilon \end{equation}
with $g$ a coupling constant and the minus partner potential is given by:
\begin{equation}V_ - ^{(1)}(x) = {g^2}{x^6} + 2\epsilon g{x^3} + {\epsilon ^2} - 3g{x^2}\end{equation}
finally the ground state wave function is identified with the first function:
\begin{equation}\psi _\epsilon ^{(1)}(x) = {e^{ - \frac{{g{x^4}}}{4} - \epsilon x}}\end{equation}
For the second function $\psi_\epsilon^{(2)}(x)$ the superpotential is:
\begin{equation}{W^{(2)}}(x) = g\sinh(x) + \epsilon \end{equation}
and the minus partner potential is:
\begin{equation}V_ - ^{(2)}(x) = {g^2}{\left( {\sinh x} \right)^2} + \epsilon^2 + 2\epsilon g\sinh x - g\cosh x\end{equation}
and the ground state wave function is:
\begin{equation}\psi _\epsilon ^{(2)}(x) = {e^{ - g\cosh (x) - \epsilon x}}\end{equation}
Finally for the third function $\psi_\epsilon^{(2)}(x)$  the superpotential is
\begin{equation}{W^{(3)}}(x) =  - \frac{{\Phi '(x)}}{{\Phi (x)}} + \epsilon \end{equation}
the minus partner potential is given by
\begin{equation}V_ - ^{(3)}(x) =  - 2\epsilon \frac{{\Phi '(x)}}{{\Phi (x)}} + {\epsilon ^2} + \frac{{\Phi ''(x)}}{{\Phi (x)}}\end{equation}
and the ground state wave function is:
\begin{equation}\psi _\epsilon ^{(2)}(x) = \Phi (x){e^{ - \epsilon x}}\end{equation}
The potentials and ground state ground state wave functions in position space are plotted in figures 20-22. The  parameter $\epsilon$ doesn't break supersymmetry but causes the potential and ground state wave function to break parity invariance $x \rightarrow -x$. The parameter $\epsilon$ affects the existence of zeros in momentum space and the Riemann hypothesis is equivalent to the statement that there are no zeros if the parameter $\epsilon$ is nonzero. Supersymmetric quantum mechanics can be realized on quantum computers where the Hamiltonian and quantum states are realized in terms of qubits \cite{Kane}
\cite{Miceli}
\cite{Apanavicius:2021yin}
\cite{Feng:2022nyi}
\cite{Culver:2023iif}
\cite{Culver:2021rxo}. The ground state wave function can be determined from the Variation Quantum Eigensolver (VQE) or Quantum Phase Estimation (QPE) which can be used in conjunction with the QFT algorithm to determine the zeros associated with the three functions above. The identification of the three functions as the ground state of the Hamiltonian has some advantages in understanding the properties of the function. For example ground state wave functions do not have nodes or zeros in position space but may or may not have them in momentum space. Also the ground state wave function may or may not be symmetric in position space but the wave function is always symmetric in momentum space.

In supersymmetric quantum mechanics one defined an operator $A$ given by:
\begin{equation}A = ip + W(x)\end{equation}
such that
\begin{equation}A\psi _\varepsilon ^{(i)} = 0\end{equation}
For the first function with $\epsilon  =0$ this operator is
\begin{equation}A = ip + g{x^3}\end{equation}
so that in position space using $p =  - i\frac{\partial }{{\partial x}}$ this becomes:
\begin{equation}\left( {\frac{\partial }{{\partial x}} + g{x^3}} \right)\psi _0^{(1)}(x) = 0\end{equation}
In momentum space using $x = i\frac{\partial }{{\partial p}}$ we have:
\begin{equation}\left( {ip - ig\frac{{{\partial ^3}}}{{\partial {p^3}}}} \right)\tilde \psi _0^{(1)}(p) = 0\end{equation} or equivalently
\begin{equation}\left( { - \frac{{{\partial ^3}}}{{\partial {p^3}}} + \frac{1}{g}p} \right)\tilde \psi _0^{(1)}(p) = 0\end{equation}
For $g=\frac{1}{64}$ we have for the generalized Airy function
\begin{equation}\left( { - \frac{{{\partial ^3}}}{{\partial {p^3}}} + 64p} \right)A{i_3}(p) = 0\end{equation}
an equation we shall see in the next section related to the $(3,1)$ Random two Matrix model.

\begin{figure}[!htb]
\centering
\minipage{0.48\textwidth}
  \includegraphics[width=\linewidth]{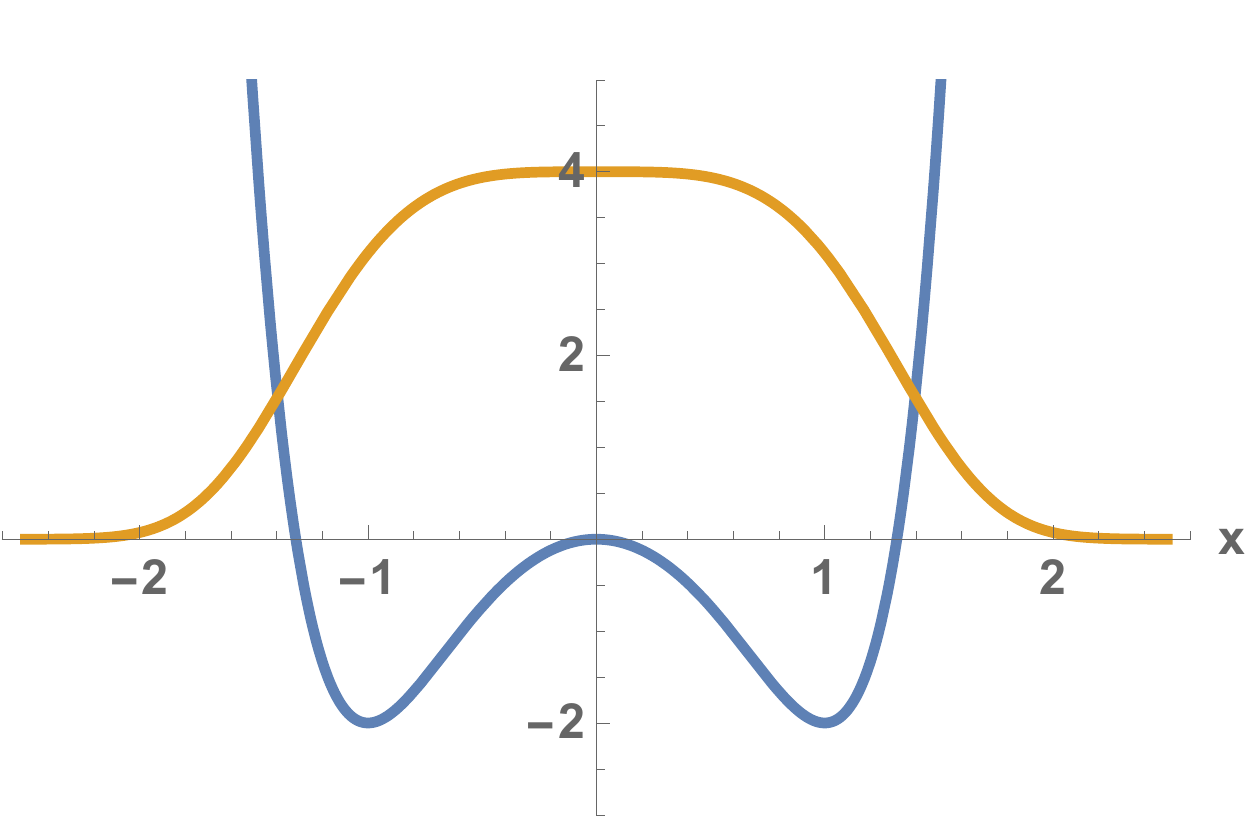}
\endminipage\hfill
\minipage{0.48\textwidth}
  \includegraphics[width=\linewidth]{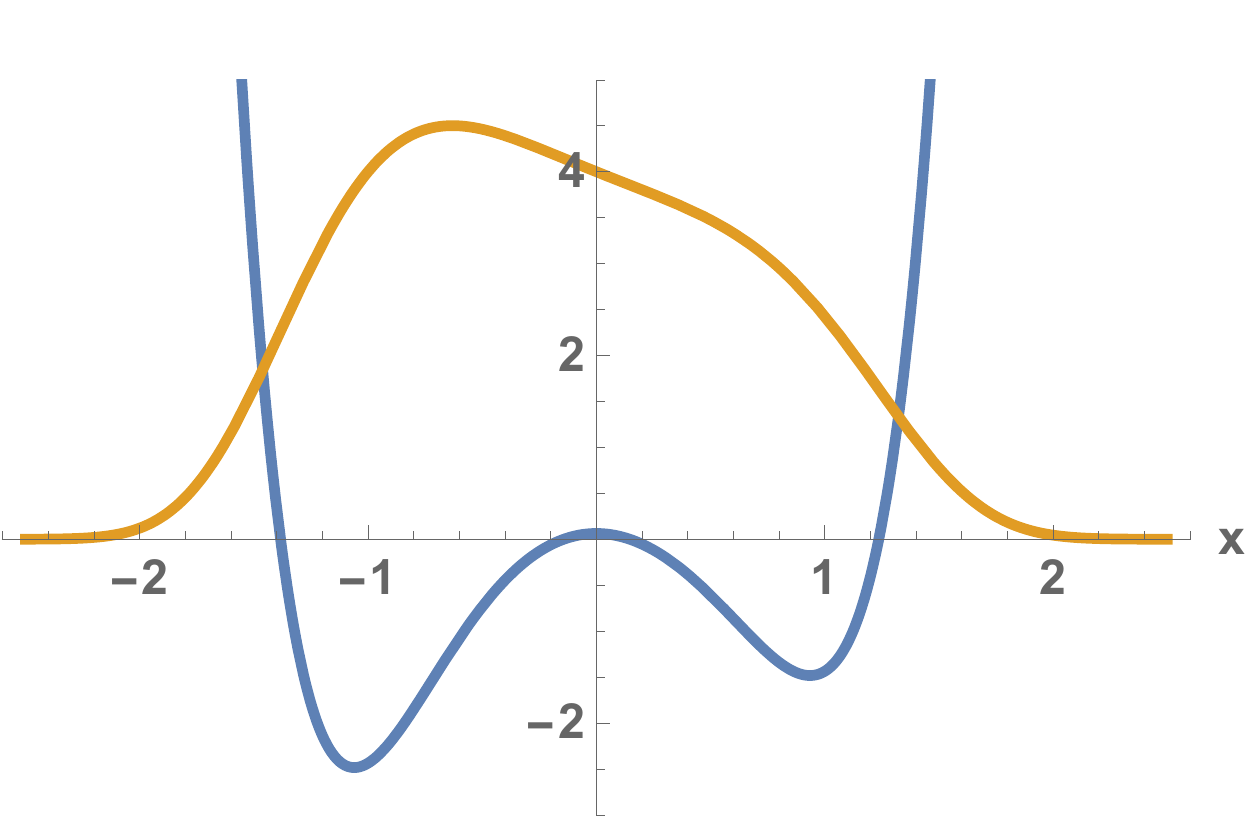}
\endminipage\hfill
\caption{ Minus partner potential (blue) and first ground state wave function (orange) for deformation parameter $\epsilon =0$ (left) and  $\epsilon = .25$ (right). 
}
\end{figure}

\begin{figure}[!htb]
\centering
\minipage{0.48\textwidth}
  \includegraphics[width=\linewidth]{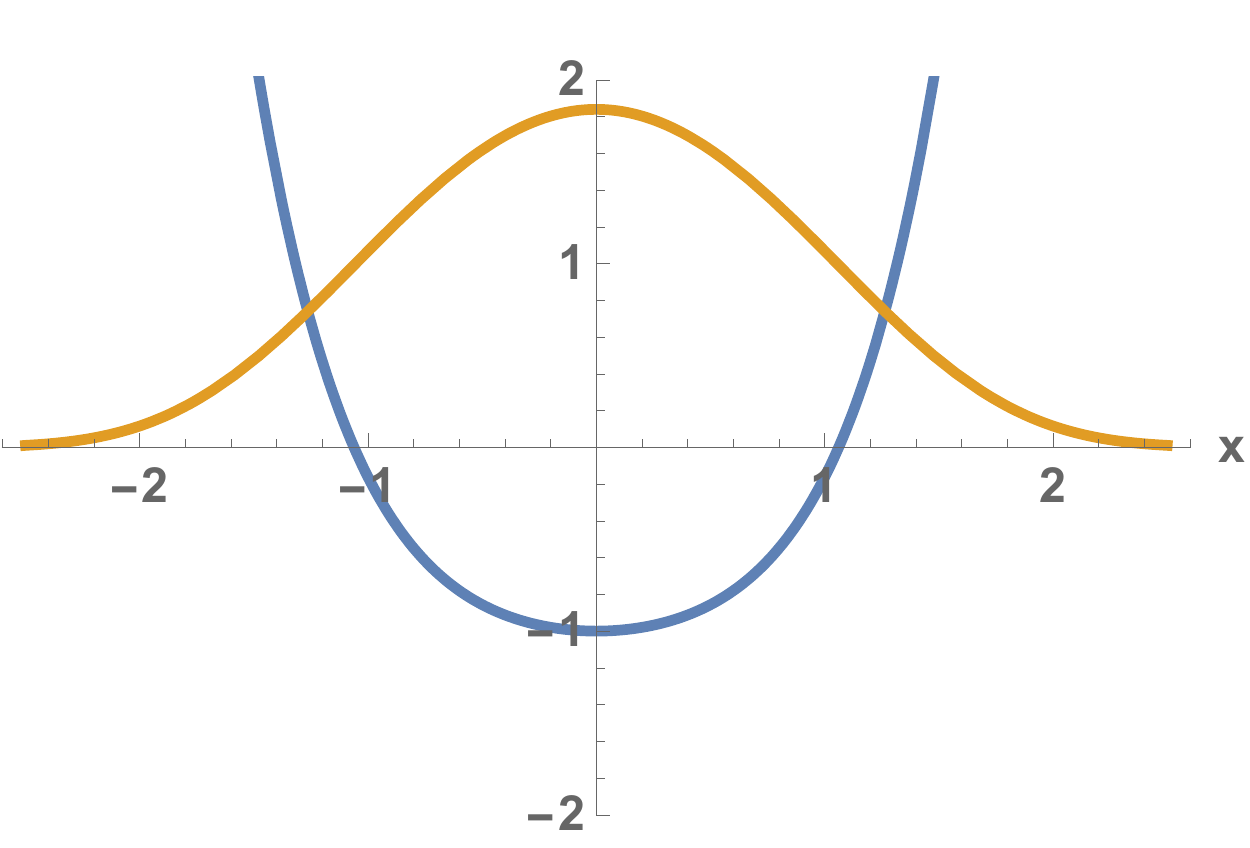}
\endminipage\hfill
\minipage{0.48\textwidth}
  \includegraphics[width=\linewidth]{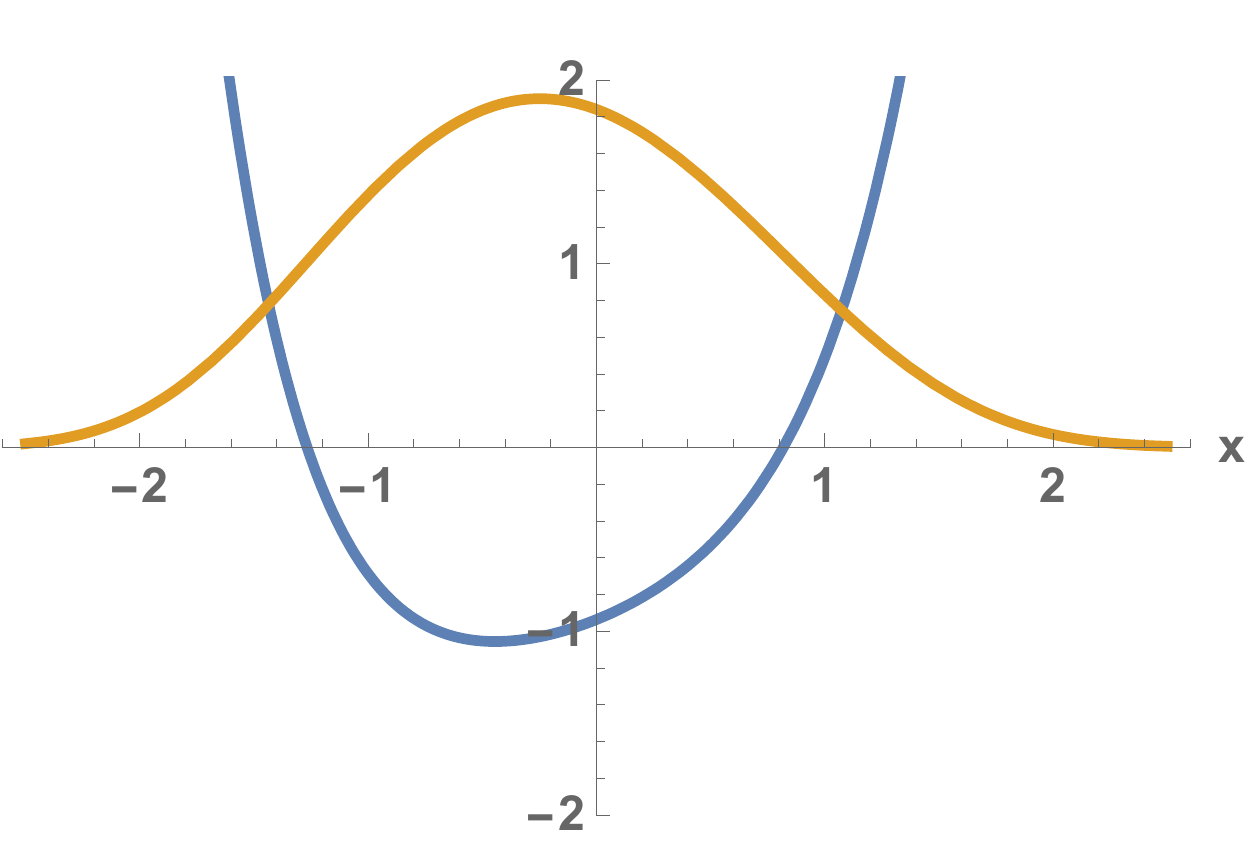}
\endminipage\hfill
\caption{Minus partner potential (blue) and second ground state wave function (orange) for deformation parameter  $\epsilon =0$ (left) and $\epsilon = .25$ (right).
}
\end{figure}

\begin{figure}[!htb]
\centering
\minipage{0.48\textwidth}
  \includegraphics[width=\linewidth]{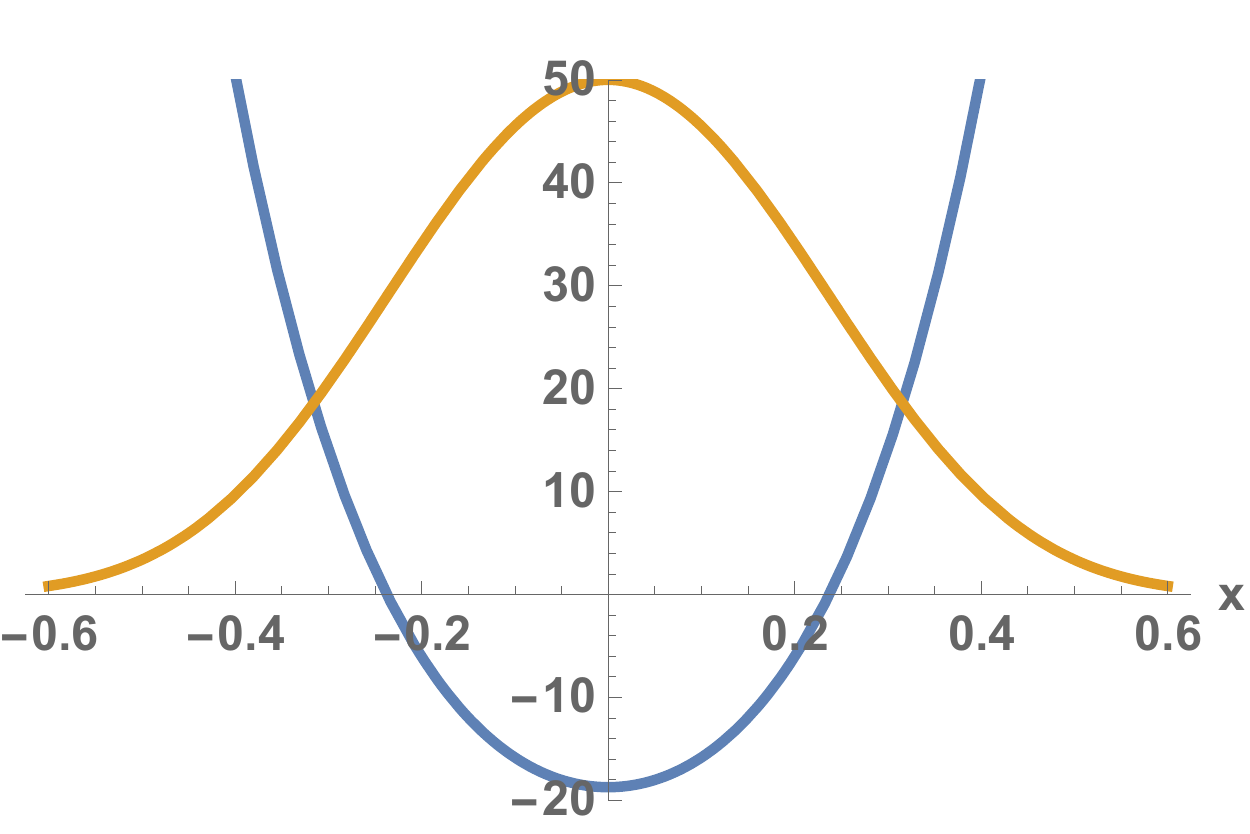}
\endminipage\hfill
\minipage{0.48\textwidth}
  \includegraphics[width=\linewidth]{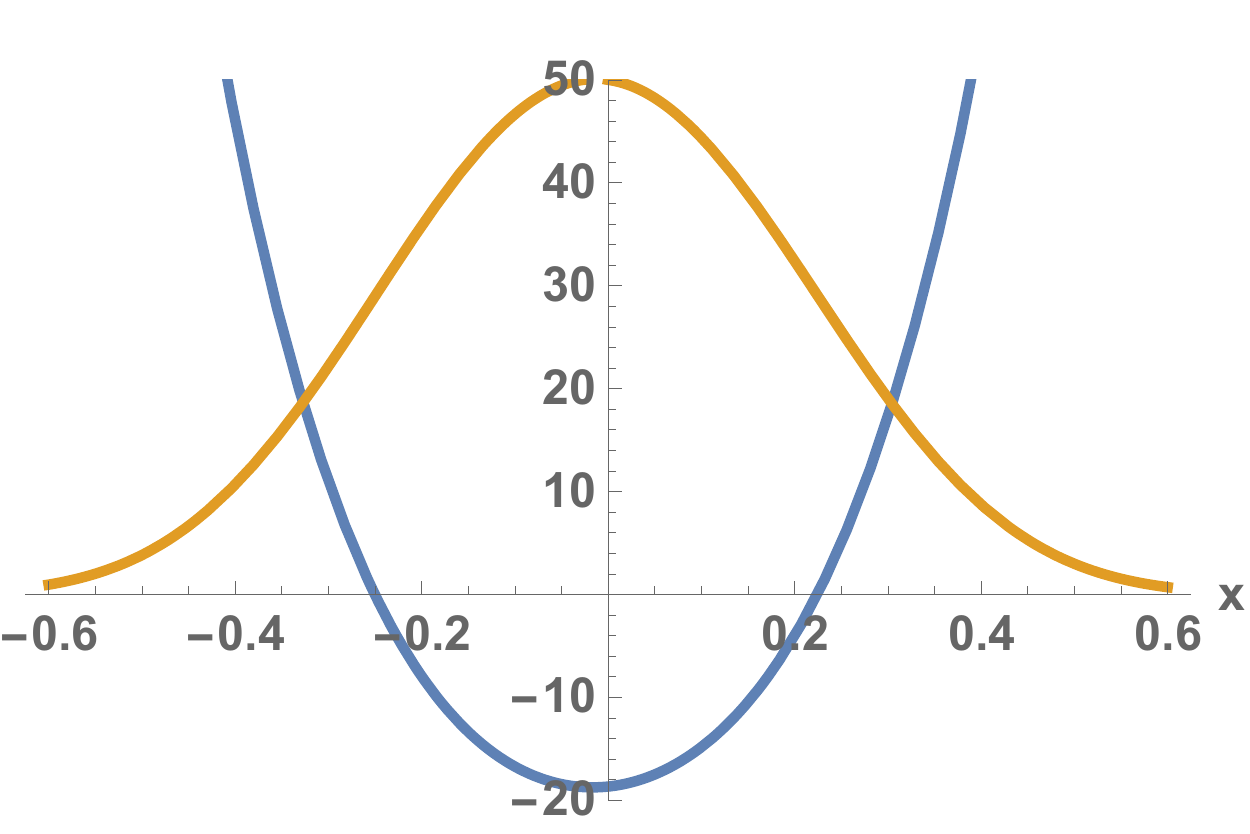}
\endminipage\hfill
\caption{Minus partner potential (blue) and third ground state wave function (orange) for deformation parameter  $\epsilon =0$ (left) and $\epsilon = .25$ (right).
}
\end{figure}

\section{Relation to Random Matrix Theory}

Random matrix theory has interesting connections to the Riemann zeta function \cite{Izenman}\cite{Tracy:1992mk}
\cite{Anninos:2020ccj}. For example the spacing of zeros on the critical line correspond to pair correlations of eigenvalues in a Gaussian Unitary Ensemble (GUE). In this section we study a connection of the type of functions in this paper to the scaling limit of a $(p,1)$ minimal Matrix model \cite{Hashimoto:2005bf}
\cite{Gopakumar:2022djw}
\cite{Gaiotto:2003yb}
\cite{Bertola:2001hq}
\cite{Ashok:2019fxv}
\cite{Seiberg:2004ei}
\cite{Seiberg:2004at}
\cite{Maldacena:2004sn}
\cite{Bleher:2002ys} where $p$ is an integer describing the order of the Matrix potential (not to be confused with $p$ used for momentum space above).  The case we will study the first function is the (3,1) Minimal Matrix model but the other two functions can be considered using  higher values of $p$ and deformations of the potential of the matrix model. The quantity of interest in relation to the zeros of the three functions is the expectation value of the characteristic potential of a hermitian matrix in the ensemble which is represented by a matrix integral as:
\begin{equation}\left\langle {\det (y - B)} \right\rangle  = {Z_{p,1}}{(g)^{ - 1}}\int {dAdB{e^{ - \frac{1}{g}Tr({V_p}(A + 1) - AB)}}\det (y - B)} \end{equation}
where $g$ is the coupling constant of the matrix model, $N\times N$ is the size of the Hermitian Matrices $A$ and $B$ where
\begin{equation}{Z_{p,1}}(g) = {\left( {2\pi g}  \right)^{{N^2}}}\end{equation} and the matrix potential is
\begin{equation}{V_p}(a) =  - \sum\limits_{k = 1}^p {\left( {\frac{{{a^k}}}{k} - \frac{1}{k}} \right)} \end{equation}
or alternatively
\begin{equation}{V_p}(a) =  - \int_1^a {\sum\limits_{k = 0}^{p - 1} {a{'^k}} da'} \end{equation}
The matrix model can be solved using orthogonal polynomials with orthonormal relations
\begin{equation}\int {dadb{e^{ - \frac{1}{g}({U_p}(a + 1) - ab)}}{P_m}(a){Q_n}(b) = {h_m}{\delta _{mn}}} \end{equation}
with $h_m$ a sequence that can be determined order by order. The solution yields the closed expression for the expectation value of the characteristic polynomial in terms of the $N$th orthogal polynomial.
\begin{equation}\left\langle {\det (y - B)} \right\rangle  = {Q_N}(y) = {\left( { - g\frac{\partial }{{\partial z}}} \right)^N}{\left. {{e^{\frac{1}{g}\left( {{U_p}(z + 1) - yz} \right)}}} \right|_{z = 0}}\end{equation}
To see the relation to our first function we specialize to the case $p=3$ so that:
\begin{equation}{V_3}(a + 1) =  - \frac{{{a^3}}}{3} - \frac{{3{a^2}}}{2} - 3a\end{equation}
It is interesting that for large $N$ the zeros of the polynomials $Q_N(y)$ are real. We plot the zeros for $N=200$ and $N=400$ in figures 23. The relation to the generalized Airy function occurs through the scaling limit where one has:
\begin{equation}g = \frac{1}{N}\end{equation}
and
\begin{equation}y = {\left( {\frac{1}{N}} \right)^{\frac{p}{{p + 1}}}}\tilde y = {N^{ - 3/4}}\tilde y\end{equation}
the generalized Airy function is given by:
\begin{equation}\psi ({N^{3/4}}y) = \frac{1}{{\sqrt {2\pi N} }}{N^{1/4}}{e^N}{e^{ - yN}}{Q_N}(y)\end{equation}
\[\psi (\tilde y) = \frac{1}{{\sqrt {2\pi N} }}{N^{1/4}}{e^N}{e^{ - \tilde y{N^{1/4}}}}{Q_N}({N^{ - 3/4}}\tilde y) = \frac{1}{{\sqrt {2\pi N} }}{N^{1/4}}{e^N}{e^{ - \tilde y{N^{1/4}}}}\left\langle {\det ({N^{ - 3/4}}\tilde y - B)} \right\rangle \]
For large $N$ the expectation value of the characteristic polynomial is the characteristic polynomial of a particular matrix called the master matrix \cite{Klinkhamer:2022pid}
\cite{Gopakumar:1994iq}
\cite{Witten}
\cite{Coleman} $B_{master}$ so that
\begin{equation}\left\langle {\det (y - B)} \right\rangle  = \det (y - {B_{master}^{(3,1)}})\end{equation}
If the master matrix is Hermitian then it's eigenvalues are real and its characteristic polynomial will have real zeros. For the $(2,1)$ matrix model the master matrix is known and is Hermitian. It is given by:
\begin{equation}B_{master}^{(2,1)} = \frac{1}{{\sqrt 2 }}\left( {\begin{array}{*{20}{c}}
0&{\sqrt 1 }&0& \ldots \\
{\sqrt 1 }&0&{\sqrt 2 }&0\\
0&{\sqrt 2 }&0& \ddots \\
 \vdots & \ddots & \ddots & \ddots 
\end{array}} \right)\end{equation}
For the $(p,1)$ matrix model the master matrix is unknown and it is not clear that if it exists it will be Hermitian and an element of the matrix ensemble or a matrix outside it. The answer depends on the infinite $N$ limit used to obtain the infinite dimensional master matrix. The Riemann hypothesis is equivalent to the statement that the master matrix exists and is Hermitian so that $\psi$ can be identified as it's characteristic polynomial. It is interesting that the Hermitian matrix $B_{master}$ of the Hilbert-Polya conjecture is not interpreted as a Hamiltonian but rather a special element of a matrix ensemble where expectation values are represented by evaluation with the master matrix. The expectation value of the characteristic polynomial is represented in the matrix model from the integration of fermions. In string field theory one can also obtain the the expectation value of the inverse of the characteristic polynomial through the integration of bosons \cite{Zeze:2015fia}:
\begin{equation}\left\langle {\frac{1}{{\det (y - B)}}} \right\rangle  = \frac{1}{{\det (y - {B_{master}})}}\end{equation}
In this case the Riemann zeros become poles and further insight may come from the interpretation in terms of scattering amplitudes of quantum fields or string theory \cite{Remmen:2021zmc}
\cite{Khuri:2001yd}. The advantage of applying quantum computing to the Random matrix model  \cite{Miceli:2019snu}
\cite{Rinaldi:2021jbg}
\cite{Chandra:2022mae}
\cite{Butt:2022xyn} is that a fault tolerant  quantum computer with a large number of qubits can examine much larger values of $N$ than a classical computer and may yield information on the large $N$ limit and the scaling behavior of the expectation value of the characteristic polynomial.

In the $(3,1)$ random two Matrix model the function $\psi(y)$ obeys the third order  differential equation \cite{Hashimoto:2005bf}
\begin{equation}\frac{{{\partial ^3}}}{{\partial {y^3}}}\psi (y) = 64y\psi (y)\end{equation}
which we identify as the equation from the previous section on supersymmetric quantum mechanics. If we denote $y_n$ as the nth zero of $\psi(y)$ we can write this equation as
\begin{equation}\frac{{{\partial ^3}}}{{\partial {y^3}}}\psi (y - {y_n}) = 64\left( {y - {y_n}} \right)\psi (y - {y_n})\end{equation}
or 
\begin{equation} - \frac{{{\partial ^3}}}{{\partial {y^3}}}\psi (y - {y_n}) + 64y\psi (y - {y_n}) = 64{y_n}\psi (y - {y_n})\end{equation}
which we recognize as an eigenproblem for a linear potential \cite{ball}
\cite{Anaya-Contreras:2020xog} with a cubic kinetic term. Returning to the momentum space representation from the previous section we write the equation as:
\begin{equation} - \frac{1}{{64}}\frac{{{\partial ^3}}}{{\partial {p^3}}}\psi _0^{(1)}(p - {p_n}) + p\psi _0^{(1)}(p - {p_n}) = {p_n}\psi_0^{(1)} (p - {p_n})\end{equation}
We plot the first seven eigenfunctions of the eigenproblem whose eigenvalues are the first seven zeros of the generalized Airy function in figure 24. Finally we note that result we obtain for the zeros from the semiclassical phase integral for the linear potential eigensystem  ${\left( {\frac{1}{4}\left( {\frac{4}{3}} \right)\left( {n + \frac{3}{4}} \right)\pi } \right)^{3/4}}$ is not as accurate for the generalized Airy function as in the usual Airy function for ${\left( {\left( {\frac{3}{2}} \right)\left( {n + \frac{3}{4}} \right)\pi } \right)^{2/3}}$  so it is important in the (3,1) Matrix model case or in the SUSY-QM treatment to use the exact expression instead of a semi-classical form.

\begin{figure}[!htb]
\centering
\minipage{0.48\textwidth}
  \includegraphics[width=\linewidth]{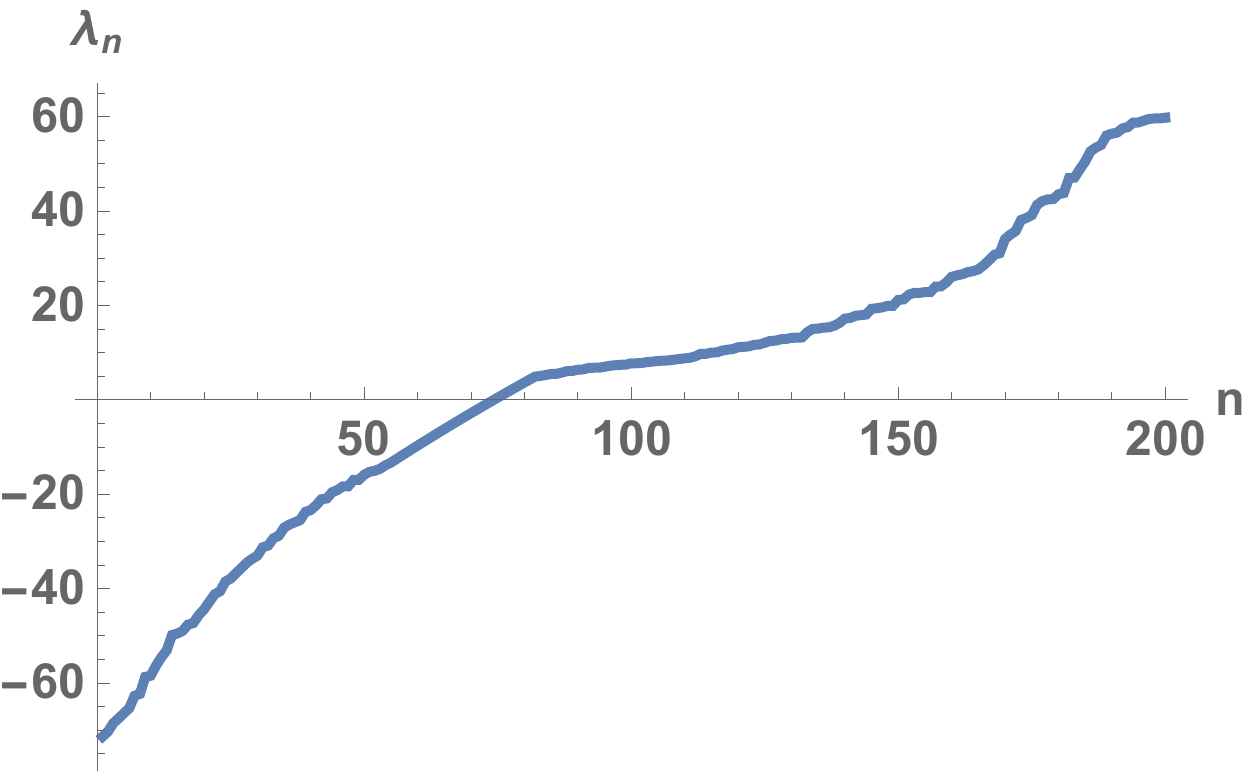}
\endminipage\hfill
\minipage{0.48\textwidth}
  \includegraphics[width=\linewidth]{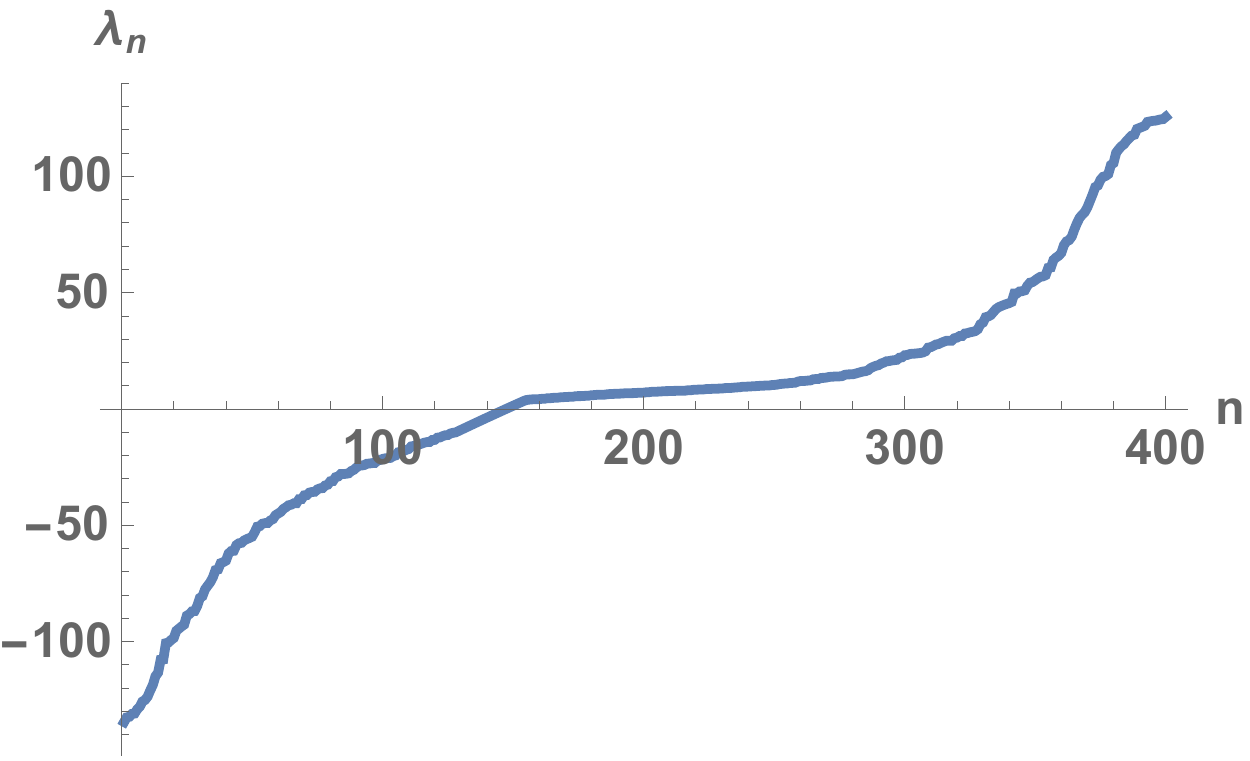}
\endminipage\hfill
\caption{Zeros of the expectation value of the characteristic polynomial for the $(3,1)$ Matrix model for (left) $N=200$ and (right) $N=400$, ordered from most negative to most positive. Surprisingly these are all real..
}
\end{figure}

\begin{figure}
\centering
  \includegraphics[width = .75 \linewidth]{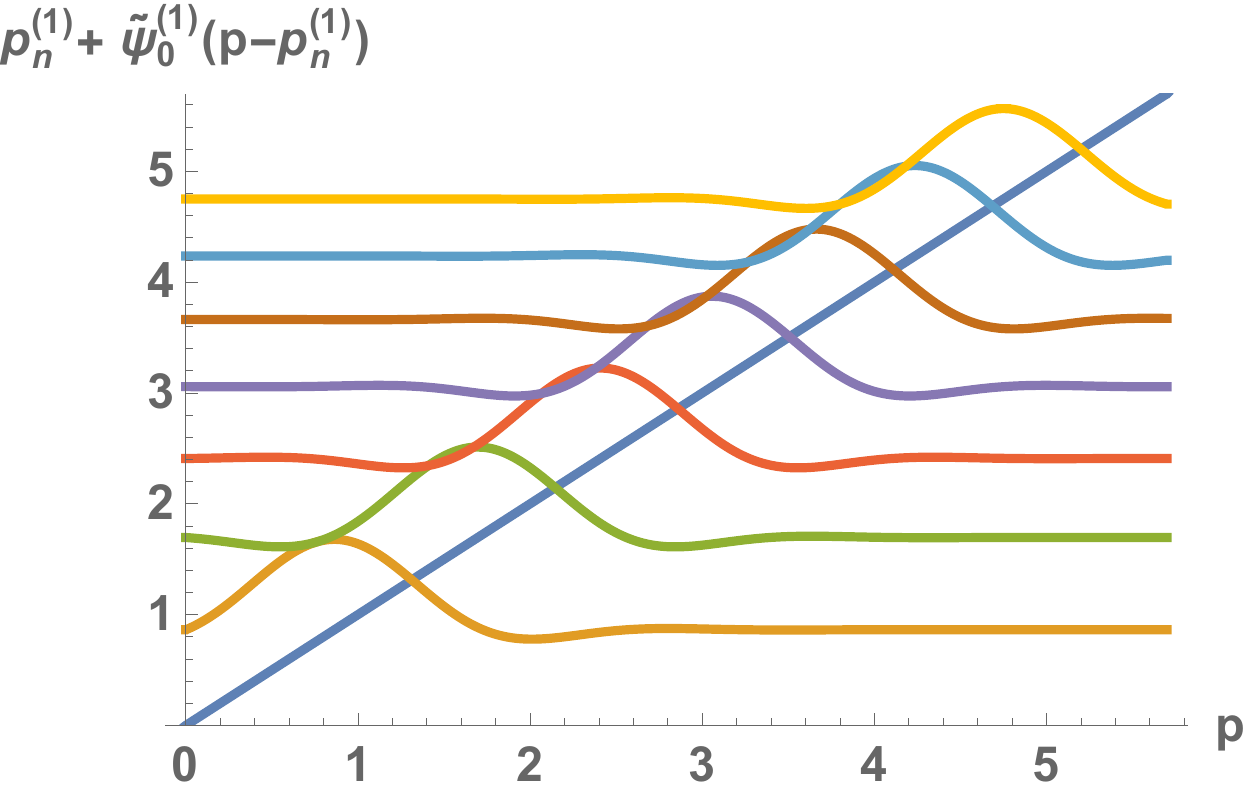}
  \caption{Solutions to the generalized Airy equation for the $(3,1)$ two matrix model and SUSY-QM with cubic superpotential transformed to momentum space and interpreted as a generalized linear potential eigensystem. }
  \label{fig:Radion Potential}
\end{figure}

\section{Conclusion}

In this paper we examined the application of quantum computing to three functions with nontrivial zeros including the Riemann xi function. We found that the quantum Fourier transform algorithm can detect zeros of the three functions using a quantum circuit and supersymmetric quantum mechanics can give a representation of the functions as a ground state of a supersymmetric quantum mechanics Hamiltonian whose spectrum can be studied on a quantum computer. Finally the representation of these functions in terms of the $(p,1)$ matrix model was studied and quantum computing can be used to examine very large matrices which can approximate the master matrix of these matrix models for large values of $N$. The intersection of quantum computing, the Riemann hypothesis, supersymmetric quantum mechanics and matrix models  is an exciting area of future research to explore. 

\section*{Acknowledgements}
We would like to to thank Yuan Feng for discussions about quantum computing applied to mathematical physics and help with the quantum computing software.

\end{document}